# The experimental lifetime α-quantization of the 36 metastable elementary particles

Malcolm H. Mac Gregor
130 Handley Street, Santa Cruz, CA 95060
Formerly at Lawrence Livermore National Laboratory,
June 1, 2008
e-mail: mhmacgreg@aol.com
URL: 70mev.org

Abstract

The fine structure constant $\alpha \cong 1/137$ appears as a lifetime scaling factor for the 36 metastable ($\tau > 10^{-21}$ sec) elementary particles—34 hadrons and the μ and τ leptons. The hadrons divide into 24 long-lived *unpaired-quark* (UQ) and 3 short-lived *paired-quark* (PQ) ground states (the *lowest-mass* states for each *u*, *d*, *s*, *c*, *b* quark flavor combination), plus 7 excited states. The lifetime ratios include six factors of α, six factors of $\alpha^4$, and one factor of $\alpha^3$ (the leptons). This experimental lifetime α-quantization has seven salient features:

(1) A *low-mass (m < 1 Gev/c*$^2$*) particle lifetime α-chain* of linked α and $\alpha^4$ lifetime ratios that extends from the neutron to the η' meson and spans 11 powers of α.

(2) Four *unpaired-quark* lifetime clusters (UQ)(π, *s*, *b*, *c*) that contain 23 ground states which are sorted by the quark priorities *c* > *b* > *s*, with the UQ clusters having *central lifetimes* $\tau_{uq}^{\pi}$, $\tau_{uq}^{s}$, $\tau_{uq}^{b}$, $\tau_{uq}^{c}$ and lifetime ratios $\tau_{uq}^{\pi}/\tau_{uq}^{s} \cong \tau_{uq}^{s}/\tau_{uq}^{b} \sim 137$ and $\tau_{uq}^{b}/\tau_{uq}^{c} \sim 3$.

(3) Four *paired-quark* clusters (PQ)(π, *s*, *b*, *c*), and four $\alpha^4$ *pairing gaps* between the *fast* PQ decays (which conserve flavor) and the *slow* UQ decays (which do not).

Superimposed on these α and $\alpha^4$ lifetime ratios are four smaller scaling effects:

(4) IR *integer-ratio scaling* inside UQ clusters, with integer lifetime ratios of 2, 3, or 4.

(5) RP *random perturbations* that produce ±10% deviations from exact IR scaling.

(6) *Hadronic* BC-*scaling*, which reflects the empirical $\mathrm{BC} \equiv \tau_b/\tau_c \sim 3$ lifetime ratios that occur between unpaired *b*-quark states and corresponding unpaired *c*-quark states.

(7) *Leptonic* BC-*scaling*, as reflected in the lifetime equation $(\tau_\tau \times \mathrm{BC})/\tau_\mu = \alpha^3$, which is accurate to 2% for BC = 3.0, and which utilizes an $\alpha^3$ scaling factor.

The UQ and PQ lifetime clusters clearly reflect their *u*, *d*, *s*, *c*, *b* quark structures.

---

## 1. Seven global features of the metastable ($\tau > 10^{-21}$ sec) elementary particle lifetimes

The experimental lifetimes of the metastable elementary particles have, almost from the beginning of the high-energy accelerator era, exhibited a global structure that suggests a dependence on the fine structure constant $\alpha = e^2/\hbar c \cong 1/137.036$. Fig. 1 (top) displays the 13 metastable lifetimes known in 1970 [1], shown plotted on an α-spaced logarithmic lifetime grid that is anchored on the $\pi^{\pm}$ meson lifetime. Fig. 1 (bottom) displays the 36 metastable lifetimes known in 2008 [2], plotted on this same lifetime α-grid. As can be seen, the 1970 α-grid accurately accommodates the 23 subsequently measured metastable lifetimes. The evolution of this metastable lifetime data base has been well-documented in the literature [3-6]. In the present paper we examine the numerical systematics of this global lifetime pattern. The experimental data on metastable lifetimes are now complete enough and accurate enough that a phenomenological assessment can be made as to the relevance of the constant α to these lifetimes.

Elementary particle *lifetimes* or *mean lives* τ have been measured with ever-increasing precision over the past few decades [1-6]. The lifetimes of the long-lived particles with observable path lengths are measured directly, whereas the lifetimes of the shorter-lived states are deduced from the resonance widths Γ of their observed energies by means of the Heisenberg relation $\tau = \hbar/\Gamma$, where Γ is the *full width* of the resonance. It is useful to denote the *metastable* excitations that have lifetimes $\tau \geq 10^{-21}$ sec (1 zeptosec) as *particles*, and to denote the excitations with *shorter* lifetimes in the range $\tau \sim 10^{-22}$ to $10^{-24}$ sec as *resonances*. The 36 metastable particles with well-measured lifetimes are displayed in detail in Fig. 2, where they are sorted into quark lifetime clusters UQ (unpaired) and PQ (paired). Because of the wide spread in lifetime values, a logarithmic representation is used, with the lifetimes expressed as exponents $x_i$ to the base $\alpha \cong 1/137$, using the $\pi^{\pm}$ meson as the reference unit lifetime.

One of the major achievements of the Standard Model has been the identification of the set *u*, *d*, *s*, *c*, *b*, *t* of fractionally-charged quarks and antiquarks, denoted collectively as quark *flavors*, which make up the strongly-interacting *hadrons*. One important feature of these quark substates is their *stabilizing* effect. Hadron excitations can be generated over a continuous range of energies, but only the energies that match the various quark configura-tions produce quasi-stable hadronic particles. The long-lived hadron *particle ground states* are the *lowest-mass* states of each of the quark combinations that occur. The ground states that contain *unpaired* quarks (with no matching antiquarks) are especially stable ($\tau > 10^{-14}$ sec), since their decays require flavor transformations from one type of quark to another, which is a violation of flavor conservation and slows the decay process.

The 36 metastable particles include 29 quark ground states, of which 11 are *unpaired-quark* and 5 are *paired-quark* meson states, and 13 are *unpaired-quark* baryon states (plus associated antiparticles with matching lifetimes). Two striking features of the lifetime patterns formed by these metastable particles are: *(1)* they map onto a global $\alpha$-spaced lifetime grid that encompasses 23 orders of magnitude; *(2)* they occur in quark clusters Q that are each dominated by a single type of quark. There are two related sets of quark clusters: *(1)* a set of four long-lived *unpaired-quark* clusters (UQ); *(2)* a matching set of four shorter-lived *paired-quark* and *excited-state* clusters (PQ) with lifetimes that are shorter by a factor of $\alpha^4$ than their long-lived counterparts.

The lifetimes of these metastable particles exhibit seven key scaling features, which we list here and illustrate below with plots of the experimental data.

(1) A *low-mass lifetime α-chain*, where $\alpha \equiv e^2/\hbar c \cong 1/137.036$ is the fine structure constant. The lifetimes of six low-mass ($m$ < 1 GeV/c$^2$) particles—$n, \mu^{\pm}, \pi^{\pm}, \pi^{o}, \eta, \eta'$—can be linked together to form a sequence of $\alpha$-quantized $\alpha$ and $\alpha^4$ lifetime ratios that spans 11 powers of $\alpha$, or about 23 orders of magnitude. This continuous lifetime $\alpha$-*chain* stands as direct experimental evidence for the $\alpha$-dependence of these lifetimes.

(2) Four *unpaired-quark* clusters UQ. The metastable particles with unpaired-quark substates occur in four separated lifetime clusters, each dominated by a characteristic quark flavor. The four UQ's are, in the order of decreasing lifetimes: UQ($\pi$) — the pseudoscalar mesons; UQ($s$) — the $s$-quark-dominated hadrons; UQ($b$) — the $b$-quark-dominated hadrons; UQ($c$) — the $c$-quark-dominated hadrons. In the case of a particle with multiple flavors, such as the $B_s = b\overline{s}$ and $B_c = b\overline{c}$ mesons, the shortest-lived flavor dominates the lifetime, so that the quark priorities in classifying the UQ clusters are $c > b > s$. The UQ($b$) particles all have essentially the same lifetime value, but the UQ($\pi$), UQ($s$) and UQ($c$) particles have narrow and *integer-quantized* spreads in lifetime values. The UQ's have clearly-delineated *central lifetimes*, which are denoted as $\tau_{uq}^{\pi}$, $\tau_{uq}^{s}$, $\tau_{uq}^{b}$, $\tau_{uq}^{c}$. The $\pi$-, $s$- and $b$- cluster central lifetimes successively decrease by factors of $\alpha^{-1} \cong 137,$ and thus map onto a lifetime α-grid that is anchored on the $\tau_{uq}^{\pi}$ central lifetime (the $\pi^{\pm}$ lifetime). But the *c*-cluster central lifetime $\tau_{uq}^{c}$ is a factor of 3 shorter than $\tau_{uq}^{b}$, and hence does not accurately fit this lifetime α-grid.

(3) Four *paired-quark* clusters PQ and associated $\alpha^4$ *pairing gaps*. These four paired-quark (and/or radiative decay) PQ's correspond to the four unpaired-quark UQ's, but have lifetimes that are a factor of $\alpha^4$ shorter, since their decays conserve flavor. The four UQ's contain a total of 23 quark ground-state particles, whereas the four PQs contain just 3 ground-state particles plus a few higher-mass excitations. The four intervening $\alpha^4$ pairing gaps contain no particle lifetimes.

Four smaller scaling effects are superimposed on this global lifetime α-grid, and they tend to obscure the overall lifetime α-scaling:

(4) IR-*scaling* (IR = integer ratios of lifetimes): This is an intra-cluster spreading of UQ lifetime values, which arises from the fact that related lifetimes within a UQ differ from one another or from the group central value $\tau_{uq}$ by integer ratios IR in factors of 2, 3, or 4.

(5) RP *random perturbations*: These observed perturbations, probably due to a variety of small effects, cause ~ ±10% deviations from precise IR intra-cluster lifetime scaling.

(6) *Hadronic* BC-*scaling* (BC = 3.0 = characteristic *b*-quark to *c*-quark lifetime ratio): The empirical BC scaling factor is the lifetime ratio $\tau_b / \tau_c = 3,$ which applies generally to the unpaired *b*-quark and corresponding unpaired *c*-quark particles. Since the *b*-quark particles accurately fit onto an α-spaced lifetime grid anchored on the $\pi^{\pm}$ lifetime (which was established before their discovery), the nearby unpaired *c*-particles do *not* fit this α-grid. Hence BC = 3.0 scaling represents a different lifetime scaling law that is overlaid on the global α-scaling. It is of interest to note that the experimental $\tau_b / \tau_c = 3.0$ *lifetime* ratio closely resembles the experimental $\Upsilon_{1S} / (J/\psi)_{1S} \equiv b\bar{b} / c\bar{c} = 3.1$ ground-state *mass* ratio.

(7) *Leptonic* BC-*scaling* and the $\alpha^3$ *quantization* of the μ and τ lepton lifetimes: The masses and lifetimes of the τ lepton and the *c*-quark D mesons are comparable, which suggests applying the hadronic BC = 3.0 lifetime scaling factor also to the τ. This leads to the lepton lifetime equation $(\tau_\tau \times \mathrm{BC}) / \tau_\mu = \alpha^3,$ which is accurate to 2.1%.

The α-dependent lifetime patterns described here are experimental regularities which are largely independent of theory, although they furnish one of the strongest arguments (together with isotopic spins and electroweak decays) for the reality of the Standard Model quark substates within these particles. This lifetime α-dependence was not predicted by existing particle theories, and it will be an interesting endeavor to incorporate it into these theories. We now use experimental lifetime data [2] to illustrate these global ground-state lifetime patterns.

## 2. The 9 low-mass metastable particles in an *α-chain* of linked α and $\alpha^4$ lifetime ratios

The 36 metastable elementary particle lifetimes are displayed in Fig. 2, where they are arrayed into five lifetime groups. The bottom group in Fig. 2 contains the lifetimes of the two lowest-mass spin 1/2 fermions — the neutron *n* and muon μ. The next group up contains the seven lifetimes of the pseudoscalar mesons — $\pi^{\pm}, \pi^{o}, \eta, \eta', K^{\pm}, K^{o}_{L}, K^{o}_{S}$, which are the lowest-mass spin 0 bosons. This group divides into an unpaired-quark cluster UQ(π) and a paired-quark cluster PQ(π). All nine of the particles in the two lowest groups in Fig. 2 have masses of less than 1 GeV/$c^2$, and are denoted here as *low-mass* (LM) metastable particles. Their long-lived and well-separated lifetimes suggest that the LM global lifetime pattern should provide interesting information. The other 27 metastable particles in Fig. 2 all have masses above 1 Gev/$c^2$. They form the three top lifetime groups in Fig. 2, each of which is dominated by a single quark flavor according to the priority rule c > b > s. Each of these three flavor groups is in turn divided into long-lived UQ lifetimes and short-lived PQ lifetimes. In the present section we discuss just the LM low-mass particles.

Fig. 3 is a logarithmic plot to the base α of the lifetimes of 6 of the LM particles— the neutron and muon, and the $\pi^{\pm}, \pi^{o}, \eta$ and $\eta'$ *non-strange* pseudoscalar mesons. As a group, these 6 particles display four interesting properties: *(a)* their lifetimes are all separated by two orders of magnitude or more; *(b)* they span the entire range of metastable lifetimes; *(c)* their low masses and spin values suggest that they correspond to very basic particle types; *(d)* they combine leptons, baryons and mesons together in one global lifetime pattern. Thus the lifetime ratios of these particles should be especially revealing. As can be seen in Fig. 3, the exponents $x_i$ to the base α of these lifetimes are all close to integer values, which suggests that the constant $\alpha^{-1} \cong 137$ is a relevant scaling factor. There is no the-ory that ties $\alpha$ to these lifetimes, so we must resort to the experimentally measured lifetime ratios to see how closely they match the value 137. (Note that an $\alpha$-spaced scaling of particle lifetimes is in factors of 1/137 if ratios of *short*-to-*long* lifetimes are used, and is in factors of 137 in the other direction. For numerical studies, it is convenient to employ $\alpha^{-1} \cong 137$ lifetime ratios, but to refer to this as a scaling in "powers of $\alpha$".)

Two characteristic lifetime ratios are displayed in Fig. 3: a very large ratio that is approximately equal to four powers of $\alpha$ (eight orders of magnitude!), and a second smaller ratio that is roughly equal to $\alpha$. The two $\alpha^4$ ratios are between the $\pi^{\pm}$ and $\pi^{o}$ mesons (which are clearly related), and between the neutron and muon (whose relationship is more obscure). Each of these two ratios is approximately equal to $(137)^{-4}$, as displayed in Fig. 3, and their averaged lifetime ratio is $(137.2)^{-4}$, which accurately matches the power-of-$\alpha$ scaling factor $\alpha^4 \cong (137.036)^{-4}$. This result seems too close to be accidental. The three $\alpha$-sized lifetime ratios in Fig. 3 show deviations from $\alpha$, but their average value is $(135.5)^{-1}$, which is close to the value $\alpha = (137.0)^{-1}$.

In addition to the six LM particle lifetimes displayed in Fig. 3, the LM metastable particles also include three *strange* K mesons, whose lifetimes are displayed in Fig. 4. The common element in Figs 3 and 4 is the $\pi^{\pm}$ lifetime, which anchors the $\alpha$-spaced lifetime grid. As Fig. 4 shows, the $K_L^o$ and $K^{\pm}$ lifetimes are separated from the $\pi^{\pm}$ lifetime by approximate factors of 2 in opposite directions. These are two examples of IR integer ratio scaling. It seems *a priori* that the $K_S^o$ lifetime should be associated with the $K_L^o$ lifetime. However, the $K^{\pm}/K_S^o$ lifetime ratio is 138.3, which matches $\alpha^{-1} \cong 137.0$ and suggests a closer relationship between the $K_S^o$ and $K^{\pm}$ lifetimes. This seems confirmed by the kaon decay modes, which are $K_L^o \rightarrow \pi\pi\pi$, $K_S^o \rightarrow \pi\pi$, and $K^{\pm} \rightarrow \pi\pi$. The factor-of-2.10 displacement of the $K_S^o$ lifetime from the $x_i = 1$ grid line in Fig. 4 mirrors the factor-of-2.10 displacement of the $K^{\pm}$ lifetime from the $x_i = 0$ grid line, and is a third example of IR ~ 2 integer lifetime ratios.

There are two key points to be noted with respect to Figs. 3 and 4: *(a)* each of the six well-separated lifetime ratios in these two figures is roughly equal either to $\alpha$ (4 examples) or to $\alpha^4$ (2 examples), and the average values measured for both of these types of intervals are very close to precise $\alpha$-scalings; *(b)* the nine particle lifetimes shown in Figs. 3 and 4 represent all of the LM low-mass ($m < 1$ GeV/c$^2$) and metastable ($\tau < 10^{-21}$ sec) elementary particles that have been discovered. Hence the universality of the $\alpha$-dependent lifetime ratios among them should be taken seriously. These $\alpha$-scaled lifetimes, together with those of the higher-mass metastable particles discussed below, are summarized in Table 1. The three IR~2 integer lifetimes ratios shown in Fig. 4 are augmented by results obtained at higher mass values, as is displayed in Fig. 9.

### 3. The global $\alpha$-spaced lifetimes of 29 metastable meson and baryon ground states

The strongly-interacting hadronic particles can be accounted-for as combinations of the *u*, *d*, *s*, *c*, *b* Standard Model quarks and their corresponding antiquarks. The massive 172.5 GeV/c$^2$ *t* quark does not play a role in the particles considered here, which all have masses below 12 GeV/c$^2$. The quark *ground states* are the lowest-mass states at which each quark combination first appears. There are 16 *meson* ground states, which are formed as quark-antiquark pairs, and can be arrayed in the form of a 5 x 5 quark-antiquark matrix. The 11 *off-diagonal* elements in this matrix are *unpaired-quark* combinations with long lifetimes ($\tau > 10^{-13}$ sec), and are displayed in Table 2. The 5 *diagonal* elements in this matrix are *paired quark-antiquark* combinations with very short lifetimes ($\tau \sim 10^{-16}$ to $10^{-22}$ sec), and are displayed in Table 3.

The *baryon* ground states are formed as *unpaired-quark* triplets. The 13 baryon ground states with measured lifetimes are displayed in Table 4. There are also matching antibaryon states, although few of these have been measured. The stable proton (*uud*)—the fundamental baryon ground state—is shown at the bottom of Table 4. The $\Sigma^o$ hyperon (*uds*) is also placed at the bottom of Table 4, since its fast radiative decay to the $\Lambda^o$ (*uds*) identifies the $\Lambda^o$ as the ground state. The free neutron (*udd*) has a 15 minute lifetime, which makes it essentially stable on a hadronic time scale. The other 12 baryon ground states in Table 4 are metastable particles with lifetimes in the range of $\tau = 10^{-10}$ sec to $10^{-14}$ sec, which is comparable to the range of the unpaired-quark meson lifetimes in Table 2. The lowest-mass charmed $\Sigma_c$ baryons might logically also be included in Table 4, but their much shorter lifetimes indicate that they in some sense represent higher excited states of the $\Lambda_c$ rather than true metastable ground states.

The 29 metastable ground states of Tables 2 - 4 are plotted together on the α-spaced lifetime grid of Fig. 5, where they are divided into mesons and baryons, and where the key quark flavors are indicated in color, in accordance with the quark lifetime dominance rule *c* > b > s. The 23 unpaired-quark meson and baryon ground states of Tables 2 and 4 appear in the $x_i = 0$ to 3 lifetime logarithm range of Fig. 5, where they sort experimentally into UQ unpaired-quark lifetime clusters, with each UQ cluster characterized by a single quark flavor. The matching PQ paired-quark meson ground states of Table 3 appear in the $x_i = 4$ to 7 lifetime range. The neutron appears at $x_i \cong -5$.

As can be seen in Fig. 5, unpaired-quark meson and baryon ground states combine together in the same UQ lifetime clusters. This indicates that *(a)* the particle lifetimes of both mesons and baryons are dictated by their quark substates, and *(b)* these *u*, *d*, *s*, *c*, *b* quark substates function with respect to lifetimes in the same manner in mesons as they do in baryons. An important conclusion to be drawn from Fig. 5 is that the same underlying mechanism which places the metastable hadrons into flavor-sorted UQ and PQ clusters also separates these clusters by powers of the constant $\alpha = e^2/\hbar c \cong 1/137$, thus creating a global lifetime α-grid that spans 12 powers of α.

**4. The 23 unpaired-quark ground states and their four UQ lifetime clusters**

The 36 *metastable* elementary particles displayed in Fig. 2 include the 27 meson and baryon *quark ground states* of Fig. 5 that are in the metastable domain (the ω and ϕ have slightly shorter lifetimes). The 23 long-lived ($\tau > 10^{-14}$ sec) *unpaired-quark* ground states of Fig. 5 (with the neutron excluded) are displayed in the lifetime plot of Fig. 6. As can be seen, they are experimentally separated into four UQ lifetime clusters that each feature a dominant quark flavor. The four clusters are: *(1)* UQ(π) at $x_i \cong 0$ (the *pseudoscalar* $\pi^{\pm}$, $K^o_L$ and $K^{\pm}$ mesons); *(2)* UQ(*s*) at $x_i \cong 1$ (the *strange* hyperons and the $K^o_S$ meson); *(3)* UQ(*b*) at $x_i \cong 2$ (the *b*-mesons and *b*-baryons); (4) UQ(*c*) at $x_i \cong 2.2$ (the *c*-mesons and *c*-baryons). The clusters UQ(π, *s*, *b*, *c*) contain 3, 7, 5 and 8 hadrons, respectively, which are listed in Table 5. These UQ clusters reveal detailed features of the lifetime spectrum.

The particles in each UQ cluster except UQ(*b*) exhibit a range of closely-spaced lifetime values. An important feature of each of these four UQ clusters is the existence of a well-delineated *central lifetime* value, $\tau_{\rm uq}$. The particles that fall in the central-lifetime band of each UQ are denoted in Fig. 6 with filled (black) symbols, and the off-central-lifetime particles are denoted with open symbols. The UQ central-lifetime particles and their average lifetime values $\tau^{\pi}_{\rm uq}$, $\tau^{s}_{\rm uq}$, $\tau^{b}_{\rm uq}$, $\tau^{c}_{\rm uq}$ are shown in Table 5. The central lifetimes $\tau_{\rm uq}$ are useful for testing the accuracy of the UQ fits to the global lifetime α-grid, and also for studying the spread of lifetime values (the IR scaling) inside a UQ cluster (Sec. 6).

The central lifetimes $\tau^{\pi}_{\rm uq}$, $\tau^{s}_{\rm uq}$ and $\tau^{b}_{\rm uq}$ form a short α-spaced chain that supplements the LM α-chain displayed in Fig. 3. The numerical values $\tau^{\pi}_{\rm uq}/\tau^{s}_{\rm uq}$ = 167.0 and $\tau^{s}_{\rm ug}/\tau^{b}_{\rm ug}$ = 104.5 in this α-chain deviate substantially from the scaling factor $\alpha^{-1} \cong 137.0$, but their average value is 135.7, which is very close to $\alpha^{-1}$. These $\tau_{\rm uq}$ α-chain results are listed together with the LM α-chain results in Table 1.

## 5. The four paired-quark PQ clusters and matching UQ-PQ $\alpha^4$ lifetime gaps

Paired quark-antiquark decays, which conserve quark flavor, proceed much more rapidly (by powers of $\alpha^4$—eight orders of magnitude!) than corresponding unpaired-quark decays. Radiative decays also conserve quark flavor, and are equally rapid. The UQ unpaired-quark clusters π, *s*, *b*, *c* (Fig. 6) that feature *slow* decays have matching PQ paired-quark and/or radiative clusters π, *s*, *b*, *c* that feature *fast* decays, as displayed in Table 5. These matching UQ and PQ clusters are identified in Fig. 2. They are also shown in Fig.7, which has the intervening $\alpha^4$ gaps plotted in parallel. In this plot, the UQ central lifetimes $\tau^{\pi}_{\rm uq}$, $\tau^{s}_{\rm uq}$, $\tau^{b}_{\rm uq}$, $\tau^{c}_{\rm uq}$ of Table 5 are each placed at $x_i = 0$ in the α-based logarithmic plot, so as to illustrate the manner in which the corresponding $\tau^{\pi}_{\rm pq}$, $\tau^{s}_{\rm pq}$, $\tau^{b}_{\rm pq}$, $\tau^{c}_{\rm pq}$ PQ cluster lifetimes appear near the $x_i = 4$ grid line. These PQ cluster lifetimes are each determined by a single particle, as listed in Table 5, although PQ(*b*) and PQ(*c*) each contain two additional excited states.

For completeness, Fig. 7 also includes the neutron-to-muon lifetime ratio that was displayed in Fig. 3 (as was the $\pi^{\pm}$-to-$\pi^{0}$ lifetime ratio). The decay modes of the neutron and muon are $n^{0} \to e^{-} + \bar{\nu}_{e} + p^{+}$ and $\mu^{-} \to e^{-} + \bar{\nu}_{e} + \nu_{\mu}$, which have generically similarities with respect to the production of an electron. But the neutron decay also involves an *n*-to-*p* quark transformation *udd* → *uud* that violates flavor conservation, and therefore places the neutron lifetime (from the present point-of-view) into the category of an unpaired-quark UQ decay. The muon decay conserves lepton number and flavor (it has no flavor), and thus is in the category of a paired-quark PQ decay, which is a factor of $\alpha^4$ faster. This, of course, assumes that the neutron and muon are *intrinsically* similar. This line of reasoning accounts for the $\alpha^4$ lifetime gap between them. The precision of the $\alpha^4$ gap value suggests that the neutron and muon are in fact related particles. But this does not explain their overall position on the global lifetime α-grid of Fig. 2, which may be linked to the decay of the charged pion into a muon, $\pi^- \to \mu^- + \bar{\nu}_\mu$.

The five measurements of $\alpha^4$ PQ-to-UQ lifetime ratios displayed in Fig.7 show considerable variations, but their average value is $(136.7)^{-4}$, which closely matches the expected value $\alpha^4 \cong (137.0)^{-4}$. The Sec. 4 and Sec. 5 UQ and PQ measurements of $S = \alpha^{-1}$ and $S^4 = \alpha^{-4}$ lifetime ratios are displayed in Table 1, together with the low-mass LM measurements of Sec. 2. The fact that the individual variations in the measured $\alpha^{-1}$ and $\alpha^{-4}$ ratios displayed in Table 1 average out to values very close to S = 137 and $S^4 = (137)^4$, respectively, and do so for all subsets of the data in Table 1, suggests that a global lifetime α-dependence is in fact the guiding factor for these metastable lifetimes.

The $\alpha^4$ gaps that occur between the UQ and PQ flavor clusters in Figs. 2 and 7 are "lifetime deserts" which are devoid of particles. These $\alpha^4$ gaps actually occur at different positions on the global lifetime grid, as displayed in Fig. 2, but the $\alpha^4$ width of the gap remains more or less constant. The α-dependence of these lifetimes seems clearly tied in some manner to an α-dependence that is inherent in the quarks themselves, and these $\alpha^4$ gaps logically provide valuable clues about the nature of the quark α-structures.

## 6. IR = 2, 3, 4 integer-ratio lifetime scaling within flavor groups

A global lifetime α-scaling in powers of $S = \alpha^{-1} = 137$ and $S^4 = \alpha^{-4} = (137)^4$ is displayed in the LM lifetime plot of Fig. 3 and the UQ and PQ flavor clusters of Figs. 6 and 7. A different and much smaller scaling emerges from the lifetime ratios observed between related particles within the UQ clusters, as displayed in detail in Fig. 6. We demonstrate here that this small-scale lifetime granularity occurs in the form of lifetime *integer ratios* (IR) in factors of 2, 3 or 4 between similar particles, with an accuracy of 10% or better.

A good way to study the spread of lifetimes within a UQ cluster is to plot each lifetime relative to the central lifetime $\tau_{uq}$ of the cluster along the horizontal axis, and then spread out the UQ lifetimes along the vertical axis so as to group the related particles together. The displacements of the *off-center* lifetimes from $\tau_{uq}$ in this plot occur in an equal-spaced manner on a logarithmic lifetime scale, which indicates that they are geometric rather than linear. The results of this procedure are displayed in Fig. 8, where it can be seen that the $(\Xi^o, \Xi^-)$, $(\Sigma^-, \Sigma^+)$ and $(D^\pm, D^o)$ pairs have approximate factor-of-2 lifetime ratios, the $(K^o_L, K^\pm)$ and $(\Xi^+_c, \Xi^o_c)$ pairs have factor-of-4 lifetime ratios, and the $(\Lambda^o, \Omega^-)$ and $(\Lambda^+_c, \Omega^o_c)$ pairs have factor-of-3 lifetime ratios. This last result seems logically related to the fact that the $\Lambda$ excitations each contain 1 *s* or *c* quark, whereas the $\Omega$ excitations each contain a total of 3 *s* and *c* quarks, with the unstable *s* and *c* quarks acting as independent decay triggers.

The lifetime scaling factors of 2, 3 and 4 displayed in Fig. 8 are not exact integers, but show deviations of several percent. We can sharpen this result by taking separate averages over the factor-of-2, -3 and -4 lifetime spacings, as displayed graphically in Fig. 9, where a few other factor-of-2 IR ratios are also included. The accuracies of the averaged ratios for IR = 2, 3 and 4 are each within 2% of being exact integers. The purpose of these calculations is not to obtain precise numerical values, but to demonstrate the *granular* structure of these lifetime deviations: the intra-group IR lifetime ratios displayed in Figs. 8 and 9 really are *integers* (at least to first order).

In the examples displayed in Figs. 8 and 9 we have taken this IR scaling one step farther than just using matched pairs by also including factor-of-2 lifetime ratios within the $(K^o_L, \pi^\pm, K^\pm)$ pseudoscalar meson triad and within the $(\Xi^+_c, \Lambda^+_c, \Xi^o_c)$ charmed baryon triad. In each case, the intermediate particle (a member of the decay chain) is located at roughly the geometric mean of the two outer particles. This yields 4 more examples of factor-of-2 lifetime ratios within the 23 hadron ground states. We have also included the low-mass $(\phi^o, \omega^o)$ vector meson ground-state pair (see Table 3 and Fig. 5), whose experimental lifetime ratio is $\tau_{\phi^o}/\tau_{\omega^o} = 1.993 \pm 0.023$ [2].

A final example of an *inferred* lifetime IR scaling factor of 2 is displayed in Fig. 4, which contains an $\alpha$-grid lifetime plot of the $K^o_L$, $K^\pm$ and $K^o_S$ mesons, shown together with the $\pi^{\pm}$ reference lifetime. As can be seen, setting the $\pi^{\pm}$ lifetime to unity gives the scalings of the $K_L^o$, $K^{\pm}$ and $K_S^o$ lifetimes as ~ 2, 1/2, and $\alpha/2$, respectively, so that the $K_S^o$ lifetime represents a factor-of-2 (or 1/2) scaling with respect to the $x_i = 1$ lifetime $\alpha$-grid abscissa. The experimental ratio of the $K^{\pm}$ and $K_S^o$ lifetimes is $138.3 \cong \alpha^{-1}$, so the $K_S^o$ appears with the $x_i \cong 1$ hyperon flavor group instead of the $x_i = 0$ pseudoscalar meson group.

In assessing the validity of this IR granular lifetime structure between related types of particles, the comprehensiveness of the IR lifetime scaling is as significant as its accuracy. Fig. 9 includes 17 of the 29 hadron ground states displayed in Tables 2-4 and Fig. 5, so the IR integer lifetime ratios are common occurrences, and not just occasional coincidences. (Also, there are no "rogue" particles that have a glaringly different scaling.) This granularity in the measured lifetimes requires a theoretical explanation, and a logical place to seek it is in a corresponding granularity in the masses of these states, which then carries over to the masses of the quarks that dictate the particle lifetimes [3].

## 7. RP random perturbations that affect exact IR integer lifetime scaling

The 23 hadron unpaired-quark ground states in the four UQ clusters of Fig. 6 include 11 central-lifetime ($\tau_{uq}$) particles and 11 off-central-lifetime (displaced) particles, plus the $\pi^{\pm}$ reference lifetime. The displaced particles in a UQ cluster exhibit integer lifetime ratios IR either with respect to related lifetimes in the UQ, or to the central UQ lifetime $\tau_{uq}$, as displayed in Figs. 8 and 9. However, these IR integer ratios are not exact, but have *random perturbations* RP that produce deviations of a few percent from exact IR scaling. Furthermore, the particles that are included in the "central lifetimes" of the UQ clusters also show RP deviations of a few percent from the $\tau_{ug}$ group-averaged values of Table 5. These deviations probably arise from a variety of small lifetime effects. Fig. 10 displays the *intrinsic* RP perturbations within the 11 *central-group* lifetimes, which are plotted on an expanded logarithmic *x*-axis grid that is in units of factor-of-2 deviations from the $\tau_{uq}$ lifetime of each UQ cluster. The average absolute value of the intrinsic RP deviations in Fig. 10 is 5.5%, which sets a lower limit on the accuracy to be expected for the lifetime IR integer ratios.

The RP deviations of the 11 *non-central* lifetimes in Fig. 6 can be ascertained by applying integer IR "correction factors" of 2, 3, or 4 to bring these lifetimes into the central-lifetime domains, and then calculating their residual RP deviations from $\tau_{ug}$. This procedure is displayed in Fig. 11, where the IR corrections factors that have been applied are shown for each particle. The average absolute RP deviation for these 11 non-central particles is 9.8%, which, as expected, is somewhat larger than the 5.5% average intrinsic RP deviations of the central lifetimes, but which shows that the RP perturbations from exact IR integer lifetime scaling are in the 10% range with respect to lifetime ratio accuracies.

We can draw three scaling conclusions from the UQ lifetime systematics displayed in Figs. 1-11: *(1)* these hadron lifetimes are grouped into UQ quark flavor clusters which are spaced by factors of $\alpha^{-1} \cong 137$ (except for the BC spacing); *(2)* lifetimes within the UQ's are spaced by IR integer ratios of 2, 3, or 4; *(3)* RP deviations from exact IR ratios are roughly 10% effects.

The overall $\alpha$-scaling of the UQ lifetime clusters can be brought out more clearly if we apply phenomenological IR "corrections" to the data in order to bring the *off-center* lifetimes into the *central-lifetime* domain for each UQ cluster. This result is displayed in Fig. 12, where it can be seen that the corrected UQ($\pi$) and UQ($s$) clusters and the uncorrected UQ($b$) cluster (which has only central lifetimes) accurately match the global lifetime $\alpha$-grid, but the UQ($c$) cluster represents a different type of lifetime scaling, which is denoted here as BC ($b$-quark to $c$-quark) scaling.

The IR-corrected ground-state lifetimes in Fig. 12 bring the scaling properties of the UQ clusters more clearly into focus, but do they constitute an "improvement" in the lifetime data set? This question is addressed empirically in Sec. 10 (Fig. 17). But first we consider BC scaling, which occurs in hadronic lifetimes (Sec. 8) and also in leptonic lifetimes (Sec. 9).

## 8. The hadronic BC *b*-quark to *c*-quark lifetime scaling factor

The $b$-quark to $c$-quark ground-state lifetime ratios BC ~ 3 stand as glaring exceptions to the $\alpha^{-1}$ ~ 137 lifetime intervals displayed in Figs. 2, 5 and 6. This is an empirical result that presently has no theoretical explanation. What we can establish here is an empirical numerical value for a hadronic BC "average correction factor" that shifts the $c$- quark lifetimes onto the global α-quantized lifetime grid at the position of the *b*-quark lifetimes. Methods for accomplishing this are summarized in Table 6.

The simplest numerical value for BC within the present UQ cluster formalism is the ratio of the UQ(*b*)/ UQ(*c*) central lifetimes, $BC = \tau_{uq}^{b} / \tau_{uq}^{c} = 3.28$ (1A of Table 6). We can also average over the ratios of individual matching lifetimes (1B of Table 6), which gives BC = 3.27 for the average of the $B^o/D^o$, $B_s/D_s$ and $\Xi_b/\Xi_c^+$ pairings, although these individual ratios show considerable variation. Finally, we can extend these calculations to include IR-corrected lifetimes of Sec. 6, which gives BC = 3.28 for the average of the $B^{\pm}/(D^{\pm} \div 2)$, $\Lambda_b^o/(\Lambda_c^+ \times 2)$ and $\Xi_b/(\Xi_c^o \times 4)$ pairings (1C of Table 6), again with considerable individual variation. (Note that the measured $\Xi_b$ lifetime is a mixture of charge states.) The important point about these various determinations of BC is not the exact numerical value that is obtained for BC, but rather the fact that BC seems to represent an *intrinsic* lifetime ratio that arises from the relative stabilities of the unpaired *b* and *c* quarks themselves. Without a theoretical framework for these lifetimes, a phenomenological value for BC is the best we can obtain.

One result that seems to be verified in 1C of Table 6 is the usefulness of employing IR-corrected lifetimes for studying global lifetime systematics. Although these IR integer-ratio "corrections" are also purely phenomenological, they serve to provide additional information about the overall systematics.

Two important theoretical questions arise in connection with the empirical hadronic BC lifetime ratio: *(1)* Why does this BC ratio have a numerical value of approximately 3? *(2)* Why is it that the UQ(*b*) unpaired-quark particle lifetimes are the ones which lie on the global α-grid of Figs. 2, 5, 6 and 12, and the UQ(*c*) unpaired-quark lifetimes are the ones which are displaced by the factor BC? These empirical facts logically serve as important guides in the creation of an α-dependent theory for these metastable ground-state decays. A third question which arises is the applicability of the BC correction factor to the τ lepton, which has both a mass and a lifetime that are comparable to the D meson masses and lifetimes? This question is dealt with in the next section.

## 9. The leptonic BC tau lifetime scaling factor

The hadronic BC scaling values shown in 1A, 1B and 1C of Table 6 are obtained from the lifetime ratios of strongly interacting mesons or baryons. The weakly interacting $\tau$ lepton would seem to be in a quite different category. However, the $\tau$ lifetime is comparable to the UQ(*c*) cluster lifetimes (Figs. 1 and 2), and the $\tau$ mass is comparable to the *c*-quark D-meson masses [2]. Thus the BC correction factor could conceivably also apply to the $\tau$ lifetime. We can test this assumption by comparing the BC-corrected $\tau$ lifetime to that of the uncorrected $\mu$ lepton. The lifetime $\alpha$-grid systematics of Fig. 2 suggests the equation $\tau_\mu/(\tau_\tau \times \mathrm{BC}) = \alpha^{-3}$. Since $\tau_\mu$, $\tau_\tau$ and $\alpha$ are all accurately measured quantities, we can use this equation to determine the leptonic value of BC, which yields BC = 2.94, as displayed in (2) of Table 6. Hence the *hadronic* and *leptonic* values for BC are in close agreement. These are all phenomenological BC determinations, and the value BC = 3.0 can be selected as a common hadronic and leptonic BC scaling factor for discussion purposes.

Fig. 13 illustrates the results of applying a BC = 3.0 lifetime correction factor to both hadrons and leptons. The five particles in the UQ(*b*) cluster (which require no corrections) are at the top of Fig. 13, where they are plotted on the global lifetime $\alpha$-grid. The reciprocal of their average lifetime ratio to that of the reference $\pi^\pm$ lifetime is $(132.1)^2$, which illustrates their $\alpha^{-2}$) quantization. The second group down in Fig. 13 is the IR-corrected UQ(*c*) *meson* cluster of Fig. 12, to which the BC = 3 correction factor has also been applied. Their average $\alpha^{-2}$-scaled lifetime ratio is $(135.9)^2$. The bottom group in Fig. 13 is the IR-corrected UQ(c) *baryon* cluster of Fig. 12 with the BC = 3 correction factor added on. Their average $\alpha^{-2}$-scaled lifetime ratio is $(142.8)^2$. These three experimental cluster lifetime ratios bracket the $\alpha$-quantized value $\alpha^{-2} \cong (137.0)^2$.

The final result displayed in Fig. 13 is the application of the BC = 3 scaling factor to the $\tau$ lepton. As can be observed in Fig. 13, this BC = 3 correction is not large enough to place the $\tau$ lifetime directly on the $x_i = 2$ grid line; it is still displaced to the right. However, as can also be observed in Fig. 13, the $\mu$ lifetime is displaced to the right of the $x_i = -1$ grid line by a similar amount. Hence the uncorrected $\mu$ lifetime and the BC-corrected $\tau$ lifetime have a lifetime ratio that very close to $\alpha^{-3}$. Numerically, this lifetime ratio is $(136.08)^3$. This is the only example of an $\alpha^3$ lifetime ratio between related metastable particles, and it emphasizes the idea that leptons and hadrons share the same global lifetime $\alpha$-grid. To illustrate the accuracy of the equation $\tau_\mu/(\tau_\tau \times BC) = \alpha^{-3}$, we set BC = 3.0 and then calculate the muon lifetime, which gives $\tau_\mu = 2.243 \times 10^{-6}$ sec. The experimental muon lifetime [2] is $\tau_\mu = 2.197 \times 10^{-6}$ sec. These two values agree to within 2.1%.

A third way of studying the constant BC is to use ADI(S, BC) minimization scans of the data, which is discussed in Sec. 10 and displayed in (3) of Table 6. The 31-particle data base that is used includes both leptons and IR-corrected hadrons, to which BC corrections are superimposed. These ADI scans establish the upper limit BC < 3.057 (Fig. 19). When taken together, the Table 6 evaluations indicate that the scaling factor BC = 3.0 is a useful consensus value to use for comparing *b*-quark and *c*-quark metastable lifetimes.

## 10. Linear ADI(S) and quadratic $\chi^2$(S) minimization plots of the scaling factor S

The lifetime $\alpha$-quantization studies carried out above were based on direct calculations of lifetime ratios S between related sets of particles, where $S \equiv (\tau_{\text{long-lived}} / \tau_{\text{short-lived}})$. We can extend these determinations of the scaling factor S by making lifetime fits to the data over a continuous range of values for S, using two different minimization techniques: *(1)* a *linear* fit using lifetime *values*; *(2)* a *quadratic* fit using lifetime *values weighted by the uncertainties in the values*.

The best-fit indicator for the *linear* fit is the *Absolute Deviation from an Integer* ADI(S) index. In this procedure, the experimental lifetimes $\tau_i$ are written in the form $\tau_i/\tau_{\pi^\pm} = S^{-x_i}$, with $\pi^\pm$ the reference lifetime and S the scaling factor. Then the lifetime logarithms $x_i$(S) are calculated for each lifetime $\tau_i$. If S is the proper scaling factor for $\tau_i$, then $x_i$(S) will be close to an integer value. The ADI(S) index is the *average value* of the *absolute* deviations $|\Delta x_i|$ from integers, as given by the equation

$$\mathrm{ADI(S)} = \frac{1}{N}\sum_{i=1}^{N} |\Delta x_i|, \quad \Delta x_i \equiv x_i - I_n, \quad I_n = \text{nearest integer},$$

where N is the number of lifetime data points. The optimum value for S is the one that minimizes ADI, which is obtained by carrying out calculations over a range of S values and plotting the function ADI(S). Since the maximum absolute deviation $|\Delta x_i|$ of an indi-vidual lifetime exponent $x_i$ from an integer value is 0.5, a *random* distribution of $x_i$'s from 0 to 0.5 will yield ADI ≅ 0.25. But if the lifetimes being tested correspond to a definite scaling factor S, then ADI(S) should show a pronounced dip at this value of S.

The best-fit indicator for the *quadratic* fit to the S-spaced lifetime grid is the chi-squared sum $\chi^2(S)$, which uses lifetime values $\tau_i(\exp)$ weighted by their experimental errors $\Delta\tau_i(\exp)$, and compares them to theoretical values $\tau_i(S)$ of the lifetimes. These are inserted into the chi-squared sum [7]

$$\chi^2(S) = \sum_i \left( \frac{\tau_i(S) - \tau_i(\exp)}{\Delta\tau_i(\exp)} \right)^2,$$

which is plotted for a range of S values, with the minimum value of $\chi^2$ determining the most accurate scaling factor S.

An ADI(S) scan was carried out for the 10-particle data set of Table 7, which are metastable particles that do not require IR or BC corrections. It yielded the ADI(S) curve shown in Fig. 14. The ADI minimum for this curve is $S_{min}$ = 135.9, which is in good agreement with the Table 1 and Fig. 13 determinations of S.

A $\chi^2(S)$ scan was also carried out for the 10-particle data set. The theoretical $\pi^o$, η and η' lifetimes $\tau_i$ were calculated as $\tau_i(S) = \tau_{\pi^\pm} S^{-x_i}$, using $x_i$ = 4, 5 and 6, respectively. The two UQ(*s*) and five UQ(*b*) lifetimes were based on the same equation, using $x_i$ = 1 and 2 for UQ(*s*) and UQ(*b*), respectively. The $\chi^2(S)$ scan gave the curve shown in Fig. 15, which has $S_{min}$ = 136.0, in close agreement with Fig. 14. Thus the ADI and $\chi^2$ procedures are equally accurate in evaluating $S_{min}$ for a set of experimental lifetime data. One difficulty with the quadratic $\chi^2$ procedure [7] is that it works well in a statistical sense only if all of the data points have roughly comparable percentage errors, which is often not the case for particle lifetime measurements [2].

ADI(S) scans of the 10-particle data set were also used to investigate the effect of using other particles than the $\pi^\pm$ as the reference lifetime. The neutron, muon, $K_L^o$ and $K^\pm$ are the other long-lived metastable particles (Fig. 2). Using these five particles as reference lifetimes gives the ADI(S) curves shown in Fig. 16, which have the following $S_{min}$ and ADI($S_{min}$) values: $K^\pm$(110.2, 0.084); μ(121.2, 0.041); *n*(130.0, 0.023); $\pi^\pm$(135.9, 0.021) $K_L^o$(158.3, 0.069). Hence if the correct lifetime scaling factor is $\alpha^{-1}$ ≅ 137.0, then the $\pi^\pm$ meson emerges clearly as the proper reference lifetime for the α-quantized global lifetime grid: the ADI minimum for the $\pi^{\pm}$ curve is at $S_{min}$ = 135.9, which is the expected position; and the value ADI($S_{min}$) = 0.021 is the lowest one displayed in Fig. 16. Since random scaling corresponds to ADI ≅ 0.250, the lifetime scaling in powers of $\alpha^{-1}$ is quite accurate.

Do IR "corrections" to the experimental lifetime data (compare Figs. 6 and 12) "improve" the data, or are they mainly "cosmetic"? As an attempt to answer this question, ADI(S) scans were carried out on the 21-particle data set of Table 7, which includes lifetimes that need IR corrections, but no lifetimes that require BC corrections. ADI(S) scans were made using both the experimental lifetimes of Fig. 6 and the IR-corrected lifetimes of Fig. 12. The results of these scans are displayed in Fig. 17. The uncorrected-data ADI curve has $S_{min}$ = 135.4 with an ADI value of 0.093, and the IR-corrected-data curve has $S_{min}$ = 135.9 with an ADI value of 0.062. (By way of comparison, the 10-particle data set, which does not require IR corrections, has $S_{min}$ = 135.9 with an ADI value of 0.021.) Thus the IR corrections lower the ADI value (a cosmetic improvement), and they also give slightly closer agreement to the scaling factor $\alpha^{-1}$ ≅ 137.0. Hence the IR corrections to the data mildly improve the phenomenology, but do not materially alter the essential features of the overall scaling in factors of S. The greatest usefulness of these IR corrections is in the challenge they offer to theorists to account for their discrete integer nature and their universality, as displayed in Fig. 9.

The final ADI(S) scans were made to investigate the effect of using the BC *b*-quark to *c*-quark correction factor discussed in Secs. 8 and 9. This requires the expanded 31-particle data set of Table 7, which includes the 9 UQ(*c*) and PQ(*c*) particles, and also the τ lepton, whose similar lifetime (Figs. 1 and 2) and mass [2] relate it to the UQ(*c*) lifetime cluster. ADI(S, BC) curves as a function of S for discrete BC values from 2.7 to 3.3 are displayed in Fig. 18. The ADI($S_{min}$) minimum stays constant at $S_{min}$ = 136.1 for BC values up to 3.0, in good agreement with the $S_{min}$ values in Figs. 14 and 15 and the IR-corrected Fig. 17. The index value ADI($S_{min}$, 3.0) = 0.061 is obtained for the 31-particle data set in Fig. 18, which agrees closely with the IR-corrected index value ADI($S_{min}$) = 0.062 for the 21-particle data set in Fig. 17 that does not involve BC corrections. Thus all of these results are in good agreement at the scaling value BC = 3.0. However, when BC is increased above 3.0 in Fig. 18, the value for $S_{min}$ decreases monotonically as a function of BC, which indicates that the BC scaling factor is *over-correcting* the *c*-quark lifetimes from the stand point of having them fit is with the other particles on the α-quantized lifetime grid. Thus BC = 3.0 represents the maximum value allowed by the ADI curves displayed in Fig. 18.

The ADI(S, BC) minimization procedure used in Fig. 18 is carried out to greater precision in Fig. 19, where a fine BC scanning mesh is used to delineate the cutoff region. As can be seen in Fig. 19, the ADI($S_{min}$, BC) minimum stays constant at $S_{min}$ = 136.08 for values of BC up to BC = 3.506, and then monotonically and linearly shifts to smaller values of $S_{min}$ for BC values of 3.507 and greater. Hence the ADI(S, BC) minimization scans yield BC < 3.507 as the maximum allowed scaling factor that can be use to "α-quantize" the c-quark and τ lepton lifetimes

One of the most interesting applications of the BC correction factor occurs when we apply it to the τ lepton, which then exhibits a very accurate $\alpha^3$ lifetime scaling ratio with respect to the μ lepton, as we discussed in Sec. 9. This is the only pair of metastable particles that exhibits an $\alpha^3$ lifetime ratio (when the τ lifetime is BC-corrected). The fact that this lifetime relationship occurs among two *leptons* makes it of special interest theoretically. Both the factor of $\alpha^3$ (which is a global lifetime α-quantization result) and the factor of BC (which comes from an empirical relationship between *b*-quark and *c*-quark states) are required, which indicates that both leptons and hadrons are enmeshed in a common α-quantized lifetime network.

## 11. Improving the elementary particle metastable lifetime data base

The growth of the number of well-measured metastable elementary particle lifetimes was demonstrated in Fig. 1, which shows 13 lifetimes in 1970 and 36 lifetimes in 2008. This number has now pretty well stabilized, and further drastic improvements are not anticipated. When we move above 11 Gev/$c^2$, the present mass limit of the metastable particles, no more particles are found until 80 Gev/$c^2$ and above, where the very-short-lived W and Z gauge bosons appear. Thus the 36 metastable particle lifetimes displayed in Fig. 2 are essentially the metastable data base that we will have for the foreseeable future.

The global features of the metastable particle lifetimes, where these lifetimes are experimentally sorted into quark-dominated clusters (Fig. 2), furnish strong evidence for the existence and relevance of the Standard Model *u*, *d*, *s*, *c*, *b* quarks. However, the α-quantization of these lifetimes has not yet found its way into the Grand Unified Theories erected around the Standard Model systematics. The Standard Model, as it is now constituted, does not include a comprehensive global method for calculating elementary particle lifetimes or masses. Thus there has not been a strong motivation to improve the accuracy of the lifetime measurements, some of which were made many years ago. The conclusion we have tried to establish in the present analyses is that metastable lifetime $\alpha$-quantization really does exist. One way to sharpen this conclusion is to improve the precision of the existing lifetime measurements. If an increase in precision leads to a more accurate level of $\alpha$-quantization, then the $\alpha$-quantization itself seems more firmly established. This is an important result, because an explanation for the $\alpha$-quantization of particle lifetimes most probably requires a reciprocal $\alpha^{-1}$-quantization of particle masses [3].

Table 8 summarizes the accuracies of the lifetime measurements for the 36 metastable particles ($\tau_\phi$ and $\tau_\omega$ are $< 10^{-21}$ sec). As can be seen, 15 of these lifetime accuracies are better than 1%, which is well within the accuracy level of this logarithmic treatment of the lifetime systematics. Another 9 are accurate to 5% or better, which is still very good. However, some of the zero-charge particles that feature radiative decays (such as the $\pi^o$ and $\Sigma^o$) have relatively large errors, and it would be of interest to see if their accuracies can be improved. Also, some of the higher-mass excitations, such as the *c*-quark and *b*-quark baryon excitations, have rather poorly-determined lifetimes.

The bottom section of Table 8 lists 13 particles whose lifetimes appear to be quite close to the metastable particle domain ($\Gamma < 0.658$ MeV), but which are known only as upper limits on the width $\Gamma$. There may be some of these states whose lifetime values can be readily improved if motivation is supplied for making good use of the increased accuracy. Hopefully the present work will supply some of that motivation.

The present lifetime phenomenology is comprehensive enough that it enables us, with the aid of the Standard Model quarks, to do an adequate job of sorting out the quark-dominated regularities in the lifetimes, as displayed for example in Fig. 2. However, it is not complete enough to enable us to have predicted how the present-day *b*-quark and *c*-quark metastable particles shown at the bottom of Fig. 1 (2008) fit in with the original lifetime $\alpha$-grid established by the 13 metastable particles shown at the top of Fig. 1 (1970). Thus the *predictive* successes of this phenomenological approach to $\alpha$-quantization have been very modest. Nevertheless, the fact that these lifetimes do form a comprehensive pattern, without the appearance of any "rogue" particles that don't fit in, is a sort of *postdiction* that should not be overlooked. The question as to whether the metastable elementary particle lifetimes do or do not depend on the fine structure constant $\alpha$ is of considerable theoretical importance. The experimental lifetime systematics displayed here may be sufficient to answer this question, and to serve as a guide for further progress in this important area of elementary particle physics.

References.

Table 1. Experimental lifetime ratios S that scale in factors of $S \cong 1/\alpha$ or $S^4 \cong 1/\alpha^4$

| $S \sim 1/\alpha \sim 137$ | $S^4 \sim 1/\alpha^4 \sim (137)^4$ |
|---|---|
| **Masses below 1 Gev/c$^2$** | |
| (1) $\tau_{\mu^\pm}/\tau_{\pi^\pm} \equiv S = 84.4$ | (7) $\tau_n/\tau_{\mu^\pm} \equiv S^4 = (141.7)^4$ |
| (2) $\tau_{\pi^o}/\tau_\eta \equiv S = 165.9$ | (8) $\tau_{\pi^\pm}/\tau_{\pi^o} \equiv S^4 = (132.7)^4$ |
| (3) $\tau_\eta/\tau_{\eta'} \equiv S = 156.2$ | |
| (4) $\tau_{K^\pm}/\tau_{K_S^o} \equiv S = 138.3$ | |
| (1-4) average is S = 136.2 | (7-8) average is $S^4 = (137.2)^4$ |
| **Masses above 1 Gev/c$^2$** | |
| (5) $\tau_{uq}^{\pi}/\tau_{uq}^{s} \equiv S = 167.0$ | (9) $\tau_{uq}^{s}/\tau_{\Sigma^o} \equiv S^4 = (214.2)^4$ |
| (6) $\tau_{uq}^{s}/\tau_{uq}^{b} \equiv S = 104.5$ | (10) $\tau_{uq}^{b}/\tau_{\Upsilon_{1S}} \equiv S^4 = (105.1)^4$ |
| | (11) $\tau_{uq}^{c}/\tau_{J/\psi_{1S}} \equiv S^4 = (89.5)^4$ |
| (5-6) average is S = 135.7 | (9-11) average is $S^4 = (136.3)^4$ |
| **Excited state to ground state ratio** | |
| | (12) $\tau_{D^\pm}/\tau_{D^{*\pm}} \equiv S^4 = (111.0)^4$ |
| **Averages over all masses** | |
| (1-6) average is S = 136.0 | (7-12) average is $S^4 = (132.4)^4$ |
| **Range of lifetimes included** | |
| $\tau_n/\tau_{\eta'} \equiv S^{11} = 2.7316\times 10^{23} = (135.08)^{11} \simeq 1/\alpha^{11}$ | |

Table 2. The 11 Standard Model UQ *unpaired-quark* metastable meson ground states, with lifetimes shown exponentially in seconds.

| Quark | $\boldsymbol{u}$ | $\boldsymbol{d}$ | $\boldsymbol{s}$ | $\boldsymbol{c}$ | $\boldsymbol{b}$ |
|---|---|---|---|---|---|
| $\boldsymbol{\bar{u}}$ | | $d\bar{u}$<br>$\pi^-(2.60-8)$ | $s\bar{u}$<br>$K^-(1.24-8)$ | $c\bar{u}$<br>$D^o(4.10-13)$ | $b\bar{u}$<br>$B^-(1.64-12)$ |
| $\boldsymbol{\bar{d}}$ | $u\bar{d}$<br>$\pi^+$ | | $(s\bar{d}, \bar{s}d)^\dagger$<br>$K_L^o(5.11-8)$ | $c\bar{d}$<br>$D^+(1.04-12)$ | $b\bar{d}$<br>$B^o(1.53-12)$ |
| $\boldsymbol{\bar{s}}$ | $u\bar{s}$<br>$K^+$ | $(\bar{s}d, s\bar{d})^\dagger$<br>$K_S^o(8.95-11)$ | | $c\bar{s}$<br>$D_S^+(5.00-13)$ | $b\bar{s}$<br>$B_S^o(1.44-12)$ |
| $\boldsymbol{\bar{c}}$ | $u\bar{c}$<br>$\bar{D}^o$ | $d\bar{c}$<br>$D^-$ | $s\bar{c}$<br>$D_S^-$ | | $b\bar{c}$<br>$B_C^-(4.60-13)$ |
| $\boldsymbol{\bar{b}}$ | $u\bar{b}$<br>$B^+$ | $d\bar{b}$<br>$\bar{B}^o$ | $s\bar{b}$<br>$\bar{B}_S^o$ | $c\bar{b}$<br>$B_C^+$ | |

†Linear combination of meson states

Table 3. The 5 Standard Model PQ *paired-quark* metastable meson ground states, with lifetimes shown exponentially in seconds.

| quark | ***u*** | ***d*** | ***s*** | ***c*** | ***b*** |
|---|---|---|---|---|---|
| $\boldsymbol{\bar{u}}$ | $(u\bar{u}, d\bar{d})^{\dagger}$ $\pi^{o}(8.40-17)$ | | | | |
| $\boldsymbol{\bar{d}}$ | | $(u\bar{u}, d\bar{d})^{\dagger\ddagger}$ $\omega^{o}(7.75-23)$ | | | |
| $\boldsymbol{\bar{s}}$ | | | $s\bar{s}^{\ddagger}$ $\phi^{o}(1.55-22)$ | | |
| $\boldsymbol{\bar{c}}$ | | | | $c\bar{c}$ $J/\psi^{o}_{1S}(7.05-21)$ | |
| $\boldsymbol{\bar{b}}$ | | | | | $b\bar{b}$ $\Upsilon^{o}_{1S}(1.22-20)$ |

†Linear combination of paired-quark states ‡Lifetime $\tau < 10^{-21}$ sec

Table 4. The 13 measured Standard Model UQ *unpaired-quark* metastable baryon ground states, with lifetimes shown exponentially in seconds.

| Quarks | State | Lifetime (sec) |
|---|---|---|
| *udd* | neutron | 885.7 |
| *uus* | $\Sigma^+$ | 8.018-11 |
| *uds* | $\Lambda^0$ | 2.631-10 |
| *dds* | $\Sigma^-$ | 1.479-10 |
| *uss* | $\Xi^0$ | 2.90-10 |
| *dss* | $\Xi^-$ | 1.631-10 |
| *sss* | $\Omega^-$ | 8.21-11 |
| *udc* | $\Lambda_c^+$ | 2.00-13 |
| *usc* | $\Xi_c^+$ | 4.42-13 |
| *dsc* | $\Xi_c^0$ | 1.12-13 |
| *ssc* | $\Omega_c^0$ | 6.90-14 |
| *udb* | $\Lambda_b^0$ | 1.409-12 |
| $\begin{pmatrix} usb \\ dsb \end{pmatrix}$ | $\Xi_b$ | 1.42-12 |
| Additional unpaired-quark baryon states | | |
| *uud* | proton | stable |
| *uds* | $\Sigma^0$ (radiative) | 7.4-20 |

Table 5. The UQ unpaired-quark and PQ paired-quark ($\pi$, $s$, $b$, $c$) lifetime flavor clusters, with central lifetimes in brackets and average central lifetime values $\tau_{uq}$ and $\tau_{pq}$ listed.

| Quark cluster | Central particles [ ] | Central lifetime (sec) |
|---|---|---|
| | Unpaired-quark flavor clusters | |
| UQ($\pi$) | $K^o_L\,[\pi^{\pm}]K^o_S$ | $\tau^{\pi}_{ug} = 2.6033\times10^{-8}$ |
| UQ($s$) | $\Lambda^o, \Xi^o[\Xi^-, \Sigma^-]\Omega^-, \Sigma^+$ | $\tau^{s}_{ug} = 1.559\times10^{-10}$ |
| UQ($b$) | $[B^{\pm}, B^o, B_s, \Xi_b, \Lambda^+_c]$ | $\tau^{b}_{ug} = 1.493\times10^{-12}$ |
| UQ($c$) | $D^{\pm}[D^o, D_s, B_c, \Xi^+_c]\Lambda^+_c, \Xi^o_c, \Omega^o_c$ | $\tau^{c}_{ug} = 4.530\times10^{-13}$ |
| | Paired-quark and radiative flavor clusters | |
| PQ($\pi$) | $[\pi^o]$ $(\eta, \eta')^{\dagger}$ | $\tau^{\pi}_{pg} = 8.4\times10^{-17}$ |
| PQ($s$) | $[\Sigma^o]$ (radiative) | $\tau^{s}_{pg} = 7.4\times10^{-20}$ |
| PQ($b$) | $[\Upsilon_{1S}]$ $(\Upsilon_{2S}, \Upsilon_{3S})^{\dagger}$ | $\tau^{b}_{pg} = 1.219\times10^{-20}$ |
| PQ($c$) | $[J/\psi_{1S}]$ $(J/\psi_{2S}, D^{*\pm})^{\dagger}$ | $\tau^{c}_{pg} = 7.047\times10^{-21}$ |

$^{\dagger}$Excited states

Table 6. Empirical values of the hadronic and leptonic BC scaling factors for *c*-quark states and the $\tau$ lepton

(1) Determinations of the *hadronic* BC scaling factor

(A) Ratio of central-lifetime averages

$$BC = \tau^{b}_{uq} / \tau^{c}_{uq} = 3.28 \text{ (see Table 5)}$$

(B) Paired lifetime ratios using uncorrected lifetimes

$$B^0 / D^0 \text{ lifetime ratio} = 3.73$$

$$B_s / D_s \text{ lifetime ratio} = 2.87$$

$$\Xi_b / \Xi_c^+ \text{ lifetime ratio} = 3.21$$

Average uncorrected ratio: $BC = (\tau_b / \tau_c)_{ave} = 3.27$

(C) Pair lifetime ratios using IR-corrected lifetimes

$$B^{\pm} / (D^{\pm} \div 2) \text{ lifetime ratio} = 3.15$$

$$\Lambda_b^0 / (\Lambda_c^+ \times 2) \text{ lifetime ratio} = 3.52$$

$$\Xi_b / (\Xi_c^o \times 4) \text{ lifetime ratio} = 3.17$$

Average IR-corrected ratio: $BC = (\tau_b / \tau_c^{corr})_{ave} = 3.28$

(2) Determination of the *leptonic* BC scaling factor

$$\tau_u / (\tau_\tau \times BC) \equiv \alpha^{-3} \rightarrow \quad BC = \tau_\mu \alpha^3 / \tau_\tau = 2.94$$

(3) Combined *hadron* and *lepton* data base: ADI(S, BC) minimization with the 31-particle IR-corrected data set

$$BC < 3.057 \text{ (see Figs. 18 and 19)}$$

CONCLUSION

The universal *hadronic* and *leptonic* BC scaling factor is, within phenomenological uncertainties, BC = 3.0.

Comments

(a) When *b*-quarks and *c*-quarks occur in the same hadronic state (as in the $B_c$ meson), the *c*-quark lifetime dominates.

(b) The correction factor BC = 3.0 applies to both the unpaired *c*-quark states and the (approximately equal mass) $\tau$ lepton.

(c) The *b*- to *c*-quark ground-state *mass* ratio $(\Upsilon_{1S})/(J/\Psi_{1S}) = 3.1$ echoes the *b*- to *c*-quark ground-state *lifetime* ratio BC = 3.0.

Table 7. The 10-, 21- and 31-particle data sets for the ADI(S) and $\chi^2(\mathrm{S})$ minimization scans

10-particle data set

$[\pi^{\pm}]$ reference lifetime; $\pi^{o}, \eta, \eta', \Xi^{-}, \Sigma^{-}, \mathrm{B}^{\pm}, \mathrm{B}^{o}, \mathrm{B}_s, \Lambda_b, \Xi_b$

(no IR or BC correction factors required)

21-particle data set

10-particle data set plus

$n, \mu, \mathrm{K}^{o}_{\mathrm{L}}, \mathrm{K}^{\pm}, \mathrm{K}^{o}_{\mathrm{S}}, \Lambda, \Xi^{o}, \Omega, \Sigma^{+}, \Sigma^{o}, \Upsilon_{1S}$

(IR corrections required, no BC corrections required)

31-particle data set

21-particle data set plus

$\mathrm{D}^{\pm}, \mathrm{D}^{o}, \mathrm{D}_s, \mathrm{B}_c, \Xi^{+}_c, \Lambda_c, \Xi^{o}_c, \Omega_c, \mathrm{J}/\psi_{1\mathrm{S}}, \tau$

(IR and BC corrections required)

Table 8. Lifetime uncertainties $\Delta\tau/\tau$ in %, where $\tau \pm \Delta\tau$ (sec) is the quoted lifetime. Also, upper limits on $\Gamma$, where $\tau = \hbar/\Gamma$.

Experimental particle lifetime uncertainties $\Delta\tau/\tau$

| | |
|---|---|
| < 1% | $n, \mu, \tau, K_L^o, \pi^\pm, K^\pm, K_S^o, \Lambda, \Sigma^+, \Sigma^-, \Xi^-, D^\pm, D^o, B^\pm, B^o, (\phi, \omega)$ |
| 1-5% | $\Xi^o, \Omega^-, D_s, J/\psi_{1S}, J/\psi_{2S}, \Lambda_c^+, B_s, \Upsilon_{1S}, \Lambda_b^o$ |
| 5-10% | $\pi^o, \eta, \eta', \Sigma^o, \Xi_c^+, \Upsilon_{2S}, \Upsilon_{3S}$ |
| >10% | $D^*(2010)^\pm, B_c^\pm, \Xi_c^o, \Omega_c^o, \Xi_b$ |

Upper limits on $\Gamma$ for narrow-width resonances

| | | | |
|---|---|---|---|
| $D^*(2007)^o$ | $\Gamma < 2.1$ MeV | $\Lambda_c(2625)^+$ | $\Gamma < 1.9$ MeV |
| $D_S^*(2112)^\pm$ | $\Gamma < 1.9$ MeV | $\Sigma_c(2455)^+$ | $\Gamma < 4.6$ MeV |
| $D_{S0}^*(2317)^\pm$ | $\Gamma < 3.8$ MeV | $\Xi_c(2645)^+$ | $\Gamma < 3.1$ MeV |
| $D_{S1}(2460)^\pm$ | $\Gamma < 3.5$ MeV | $\Xi_c(2645)^o$ | $\Gamma < 5.5$ MeV |
| $D_{S1}(2536)^\pm$ | $\Gamma < 2.3$ MeV | $\Xi_c(2815)^+$ | $\Gamma < 3.5$ MeV |
| $\chi(3872)$ | $\Gamma < 2.3$ MeV | $\Xi_c(2815)^o$ | $\Gamma < 6.5$ MeV |

$\Xi_{cc}(3518.9)$ $\tau < 33\times10^{-15}$ sec (controversial state)

Note: $\Gamma$ = 0.658 MeV corresponds to $\tau = 10^{-21}$ sec

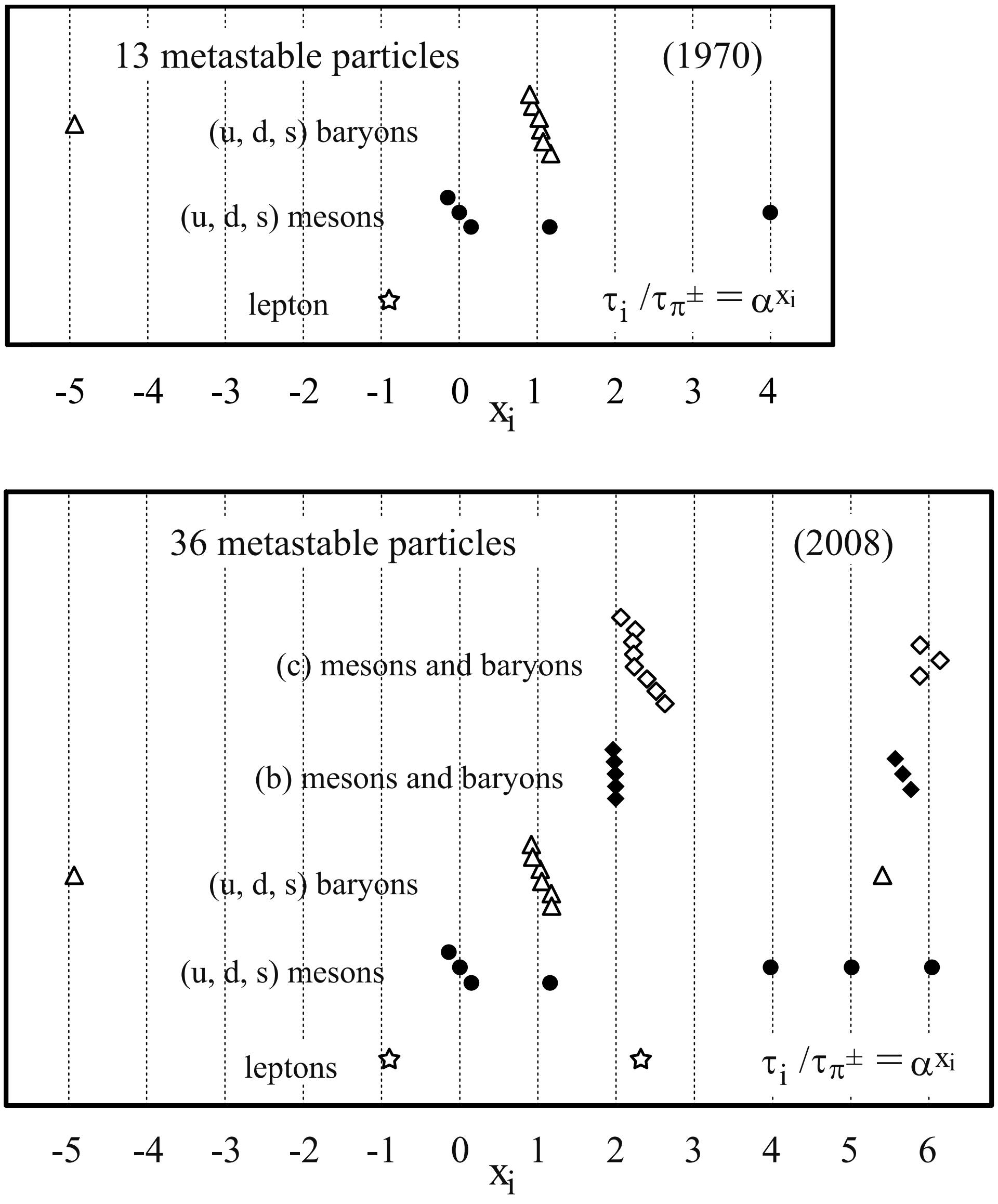

Fig. 1. The metastable ($\tau > 10^{-21}$ sec) elementary particle data base as it appeared in 1970 (13 particles), and as it appears in 2008 (36 particles). The original 1970 $\alpha$-quantized global lifetime grid still applies 38 years later.

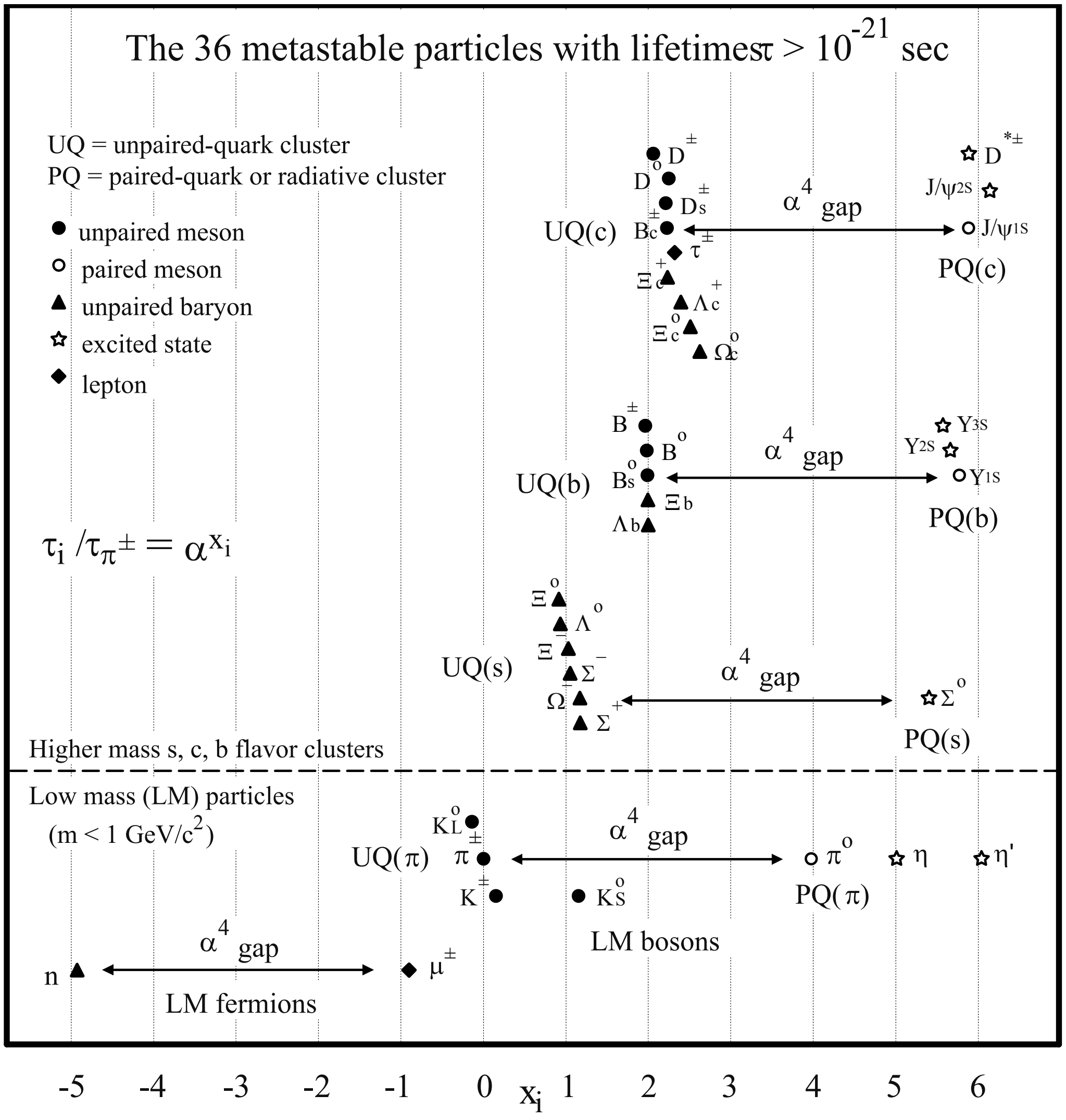

Fig. 2. The 36 metastable particle lifetimes, shown divided into five lifetime groups and plotted on an α-quantized global lifetime grid that is anchored on the $\pi^{\pm}$ lifetime. Each group in turn divides into a long-lived unpaired-quark UQ component and a much-shorter-lived paired-quark PQ component, which are separated by an $\alpha^4$ lifetime gap. The high-mass ($m > 1$ GeV/c$^2$) UQ and PQ clusters are each dominated by a single *s*, *b*, or *c* quark flavor according to the priority rule $c > b > s$. The LM low-mass ($m < 1$ GeV/c$^2$) bosons and fermions contain just the unflavored *u* and *d* quarks and the *s* quark. The μ and τ leptons fit into this same lifetime systematics.

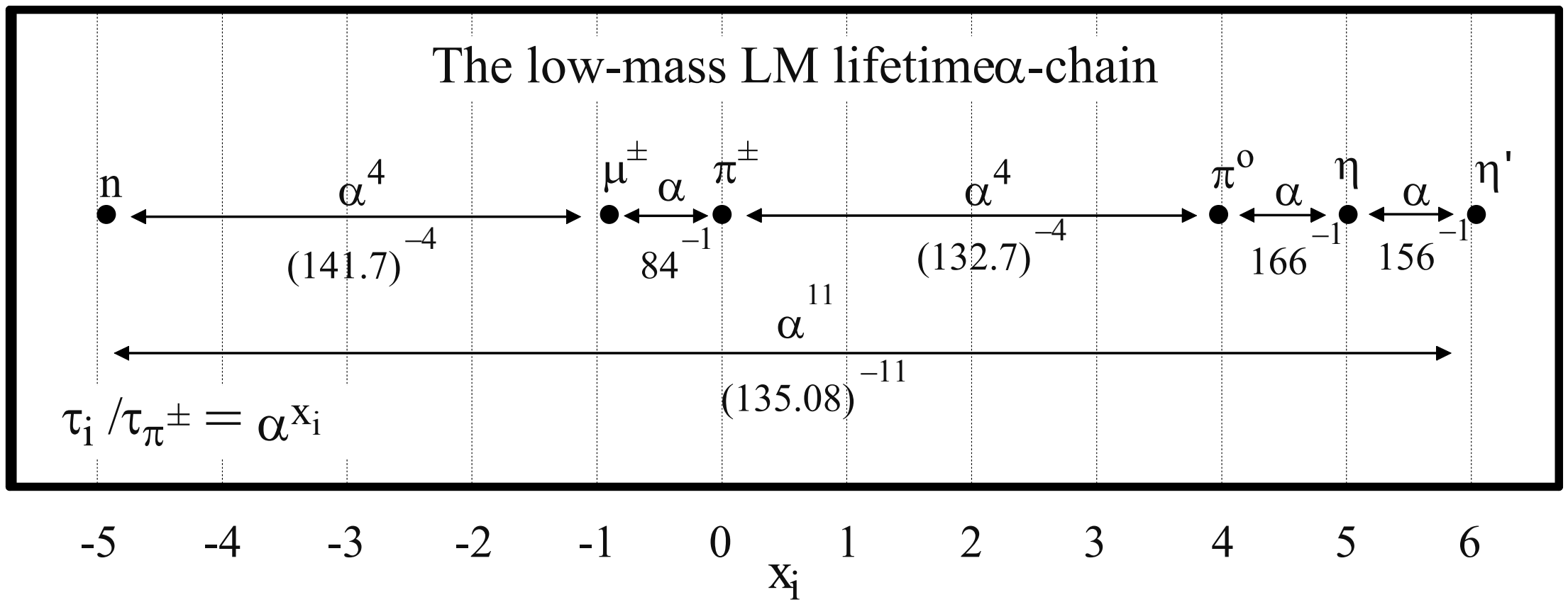

Fig. 3. An α-quantized linked chain of low-mass non-strange metastable particle lifetime ratios, which extends from the neutron to the η' meson and spans the entire metastable lifetime range (11 powers of α). The two large $\alpha^4$ links have an average lifetime ratio of $(137.2)^{-4}$, and the three small α links have an average lifetime ratio of $(135.5)^{-1}$ (Table 1). These average values reflect the theoretical $\alpha \cong 137^{-1}$ spacing of the global lifetime grid, and the experimental $(135.08)^{-11}$ total lifetime span of the linked α-chain. This low-mass lifetime chain includes a lepton, a baryon, and four unflavored mesons, all linked together in a comprehensive α-quantized global lifetime pattern.

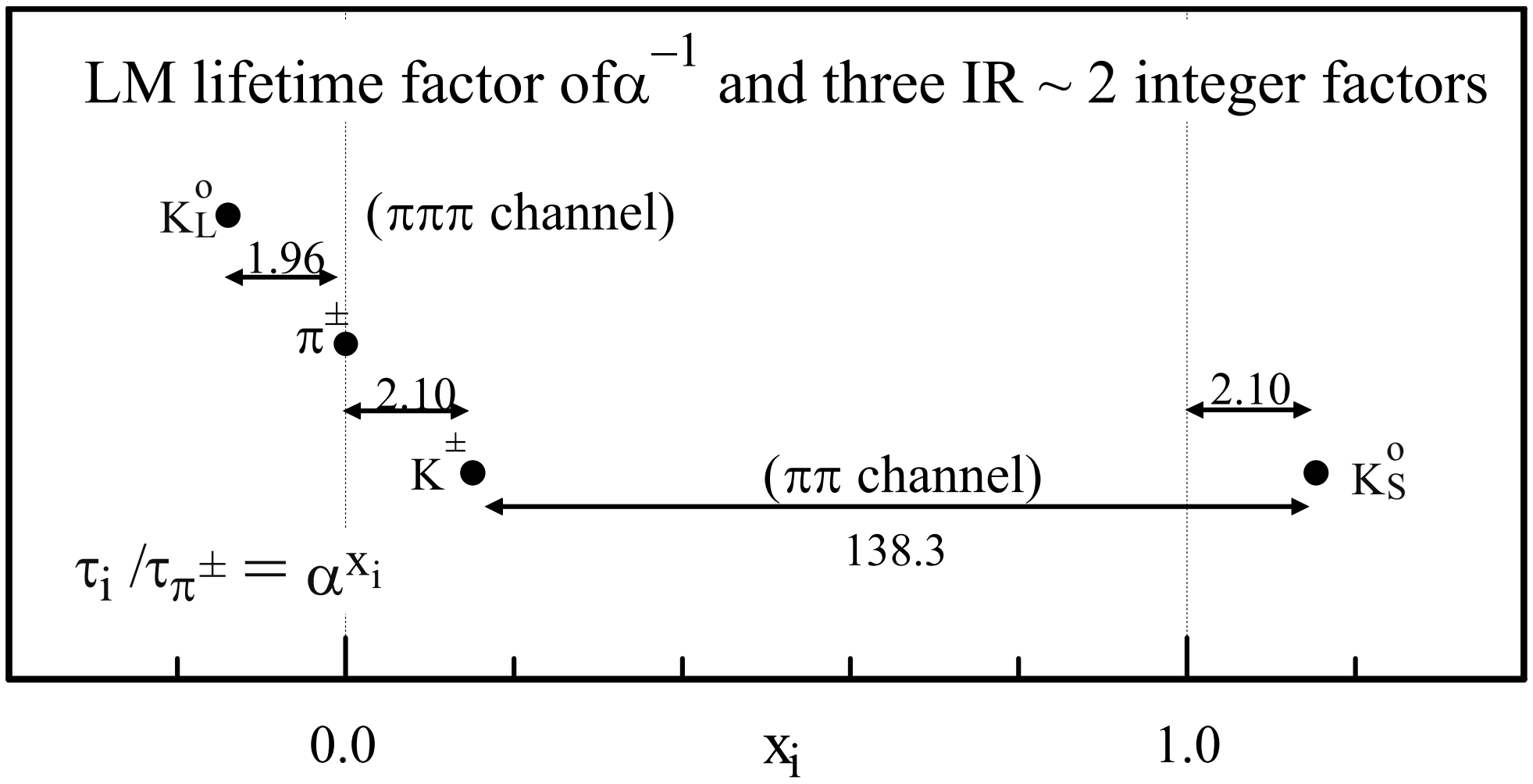

Fig. 4. The three low-mass strange meson lifetimes, shown plotted together with the reference $\pi^{\pm}$ lifetime on the α-quantized global lifetime grid. The $K^{\pm}$ and $K_S^o$ lifetimes have a very accurate α-spaced lifetime ratio, and the three kaons have quite accurate factor-of-2 IR (integer ratio) displacements from the lifetime α-grid. Other examples of these IR lifetime ratios are displayed in Figs. 8 and 9.

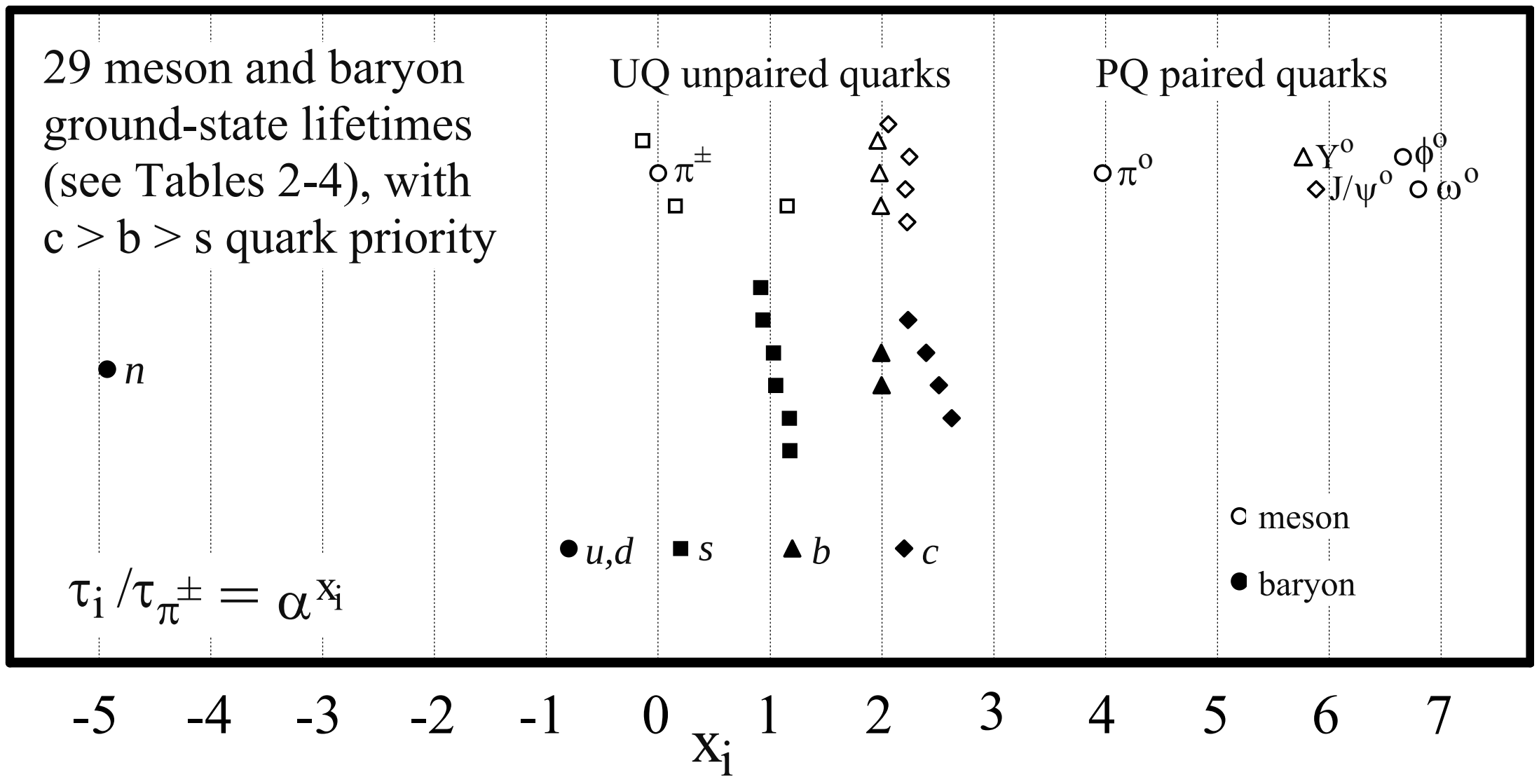

Fig. 5. The meson and baryon ground states (the lowest-mass states at which the various quark combinations first occur), shown plotted on an α-spaced global lifetime grid. The dominant quarks are labeled graphically to show how the quarks group together in flavor clusters. The mesons and baryons are separated vertically on the plot, which illustrates how they share the same quark lifetime characteristics.

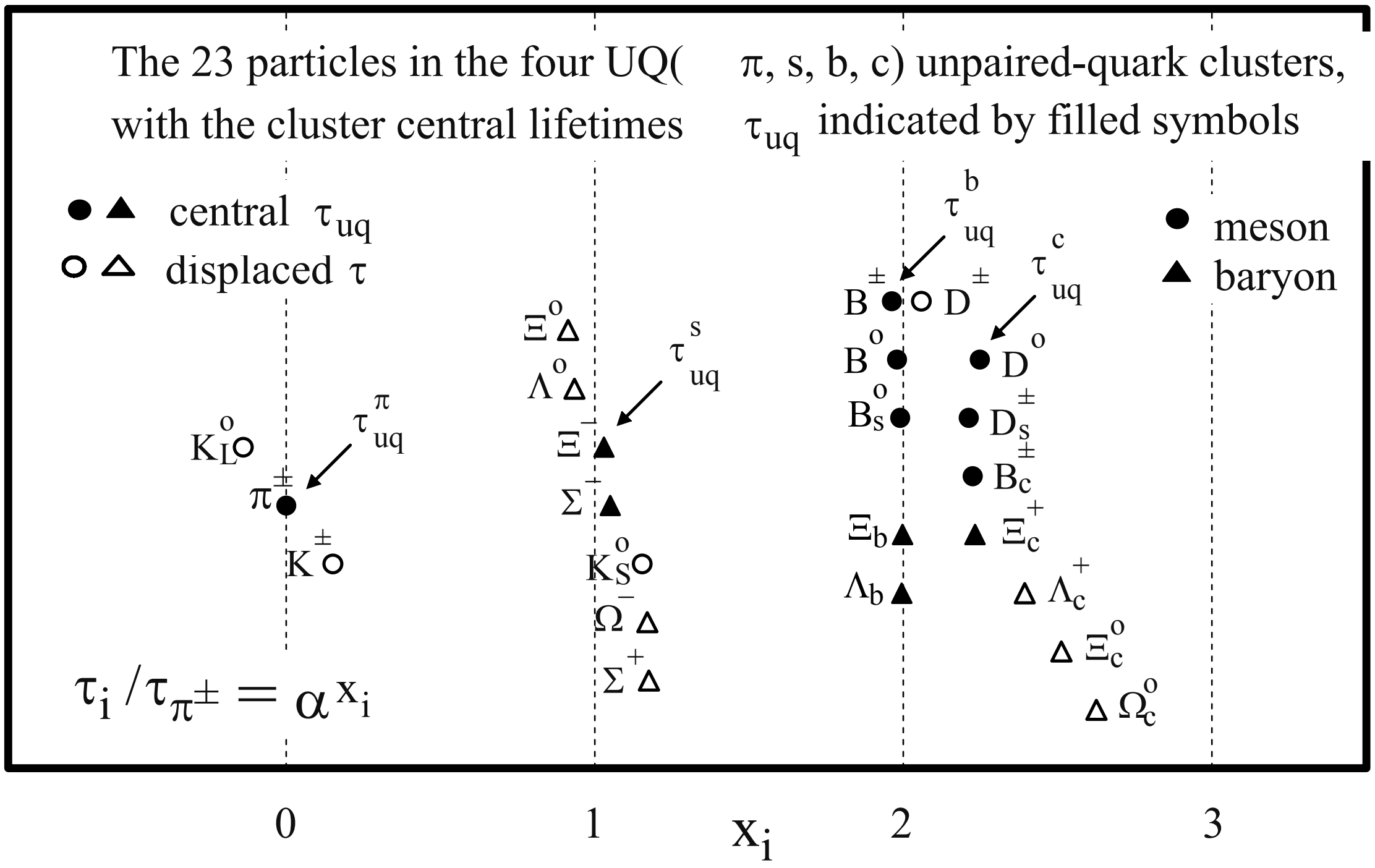

Fig. 6. The UQ unpaired-quark clusters, showing the 12 central lifetimes (solid symbols) and 11 displaced lifetimes (open symbols). The average value of the central lifetime or group of lifetimes in each cluster is designated as $\tau_{uq}$ (Table 5). The $\tau^{\pi}_{uq}$, $\tau^{s}_{uq}$ and $\tau^{b}_{uq}$ central lifetimes lie on or close to the global α-spaced vertical grid lines, but the $\tau^{c}_{uq}$ central lifetime is shifted off the α-grid with respect to $\tau^{b}_{uq}$ by the factor BC ≅ 3 (Table 6). The 11 displaced lifetimes in Fig. 6 all form fairly accurate IR integer ratios of 2, 3 or 4 with respect to $\tau_{uq}$ or to one another (Figs. 8 and 9).

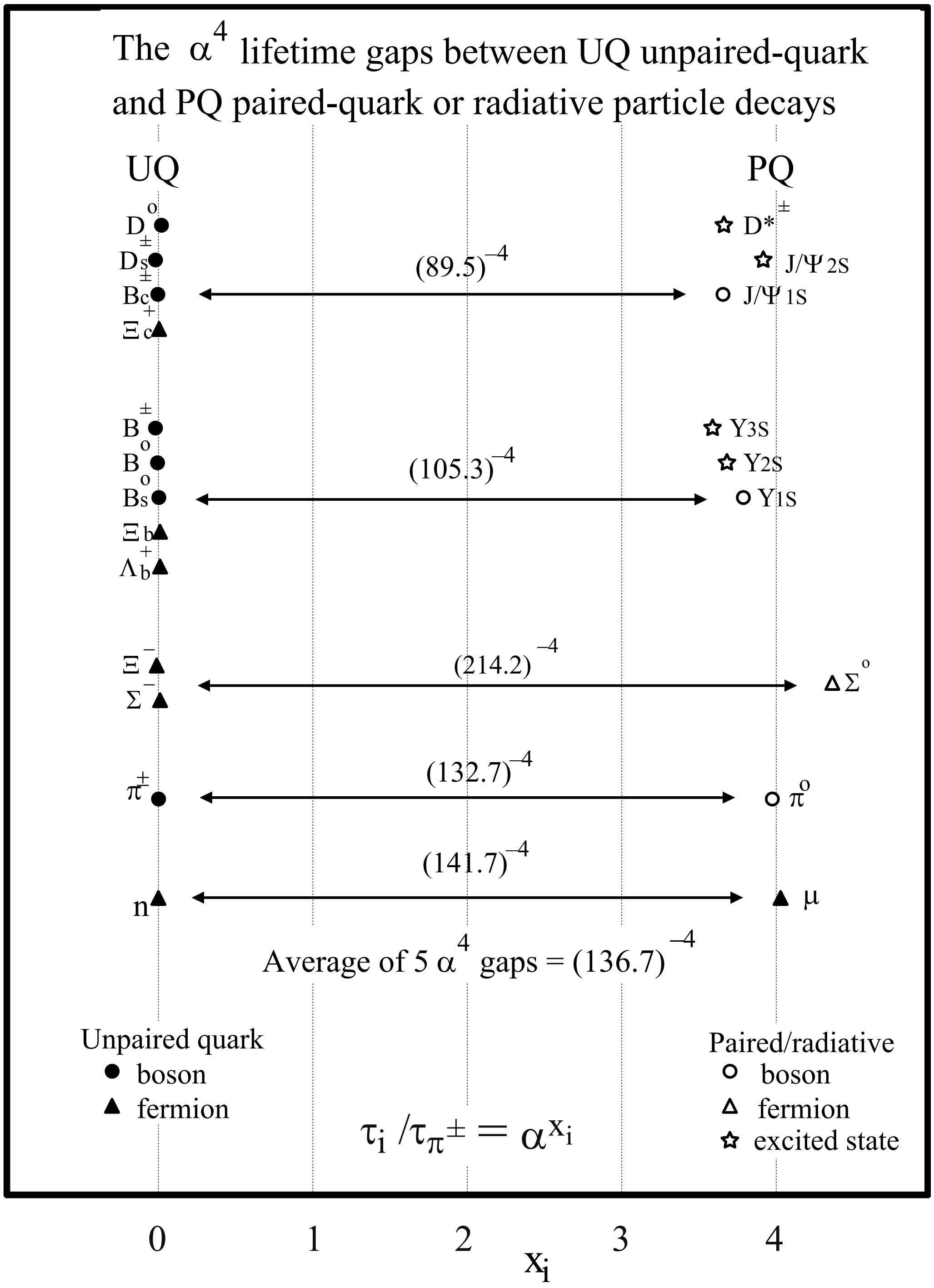

Fig. 7. The five $\alpha^4$ lifetime gaps of Fig. 2, shown with the UQ central lifetime averages (Table 5) used as the $x_i = 0$ points on a logarithmic α-scaled lifetime grid. Although these $\alpha^4$gaps have varying lengths, their average value is close to the scaling value $\alpha^4 = (137.0)^{-4}$.

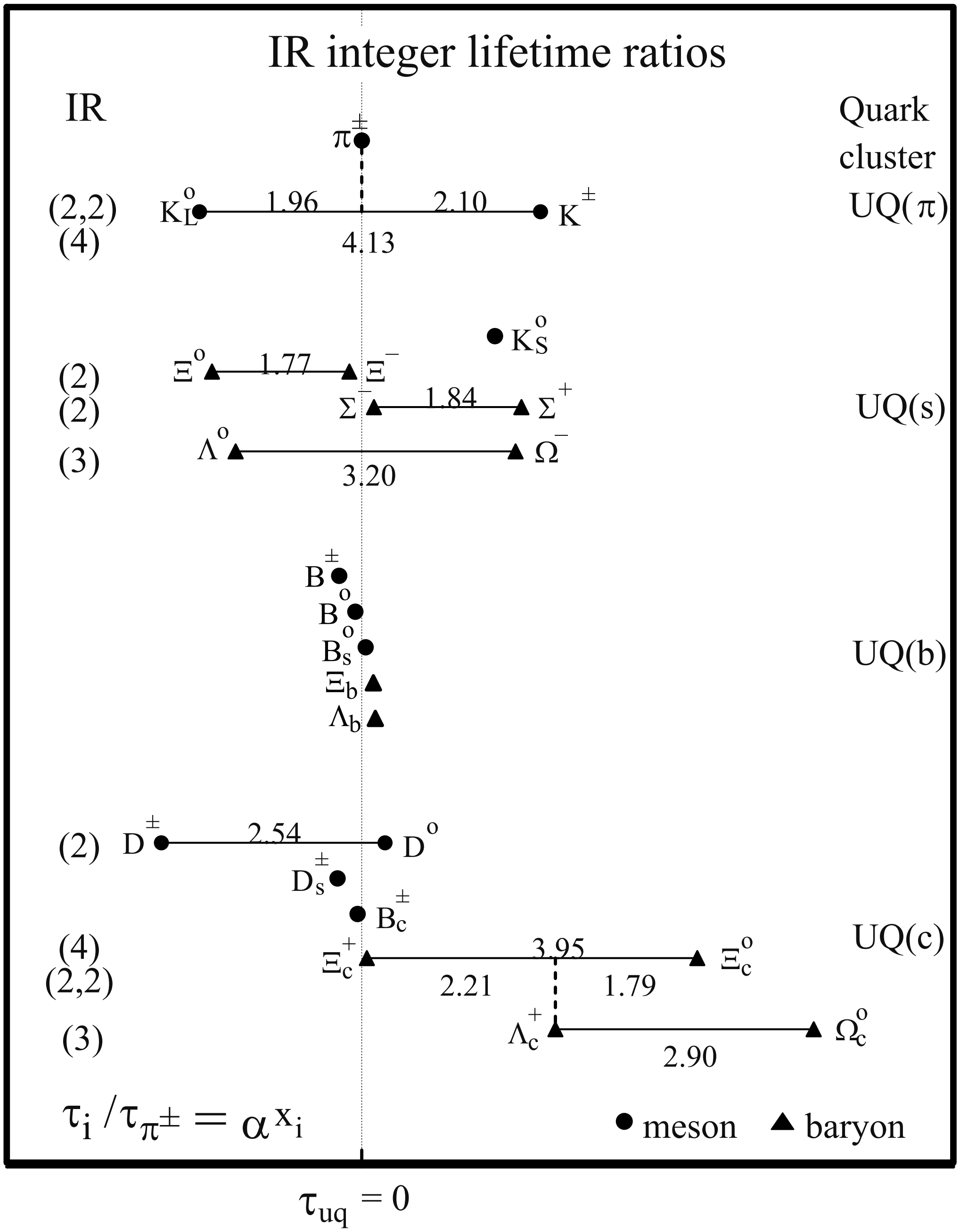

Fig 8. Intra-cluster IR integer ratios of 2, 3 and 4 for the 11 displaced lifetimes in the UQ unpaired-quark clusters of Fig. 6, with the central lifetimes $\tau_{uq}$ set at $x_i = 0$. As can be seen, these IR ratios vary somewhat from exact integers, but their average values (Fig. 9) are quite close to integers.

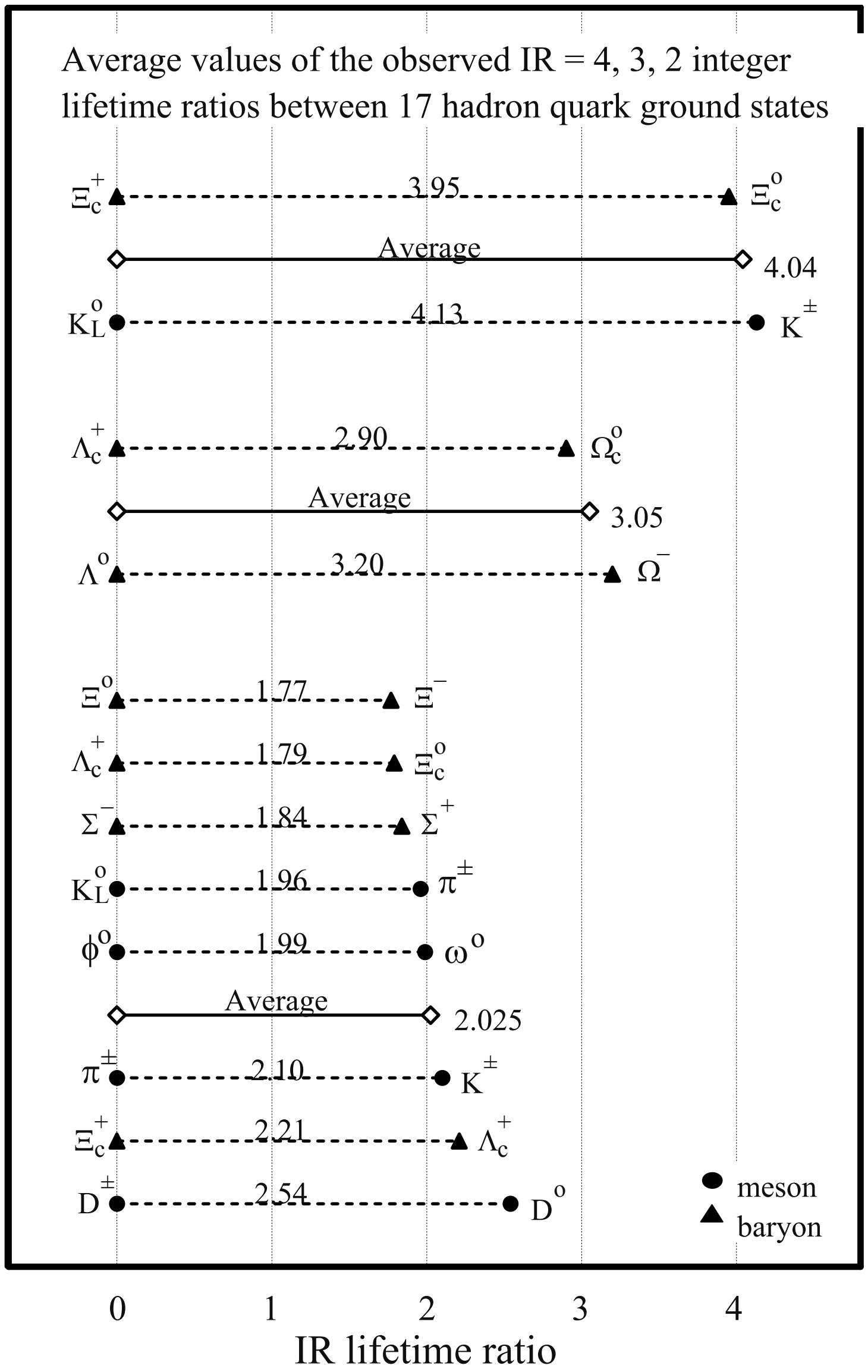

Fig. 9. The IR = 4, 3 and 2 ratios of Fig. 8, sorted out and grouped together here on a linear lifetime plot. Also included is the accurate $\phi^o/\omega^o = 1.993$ integer lifetime ratio. These experimental ratios are arrayed around their IR average values, which are each within 2% of being an exact integer value.

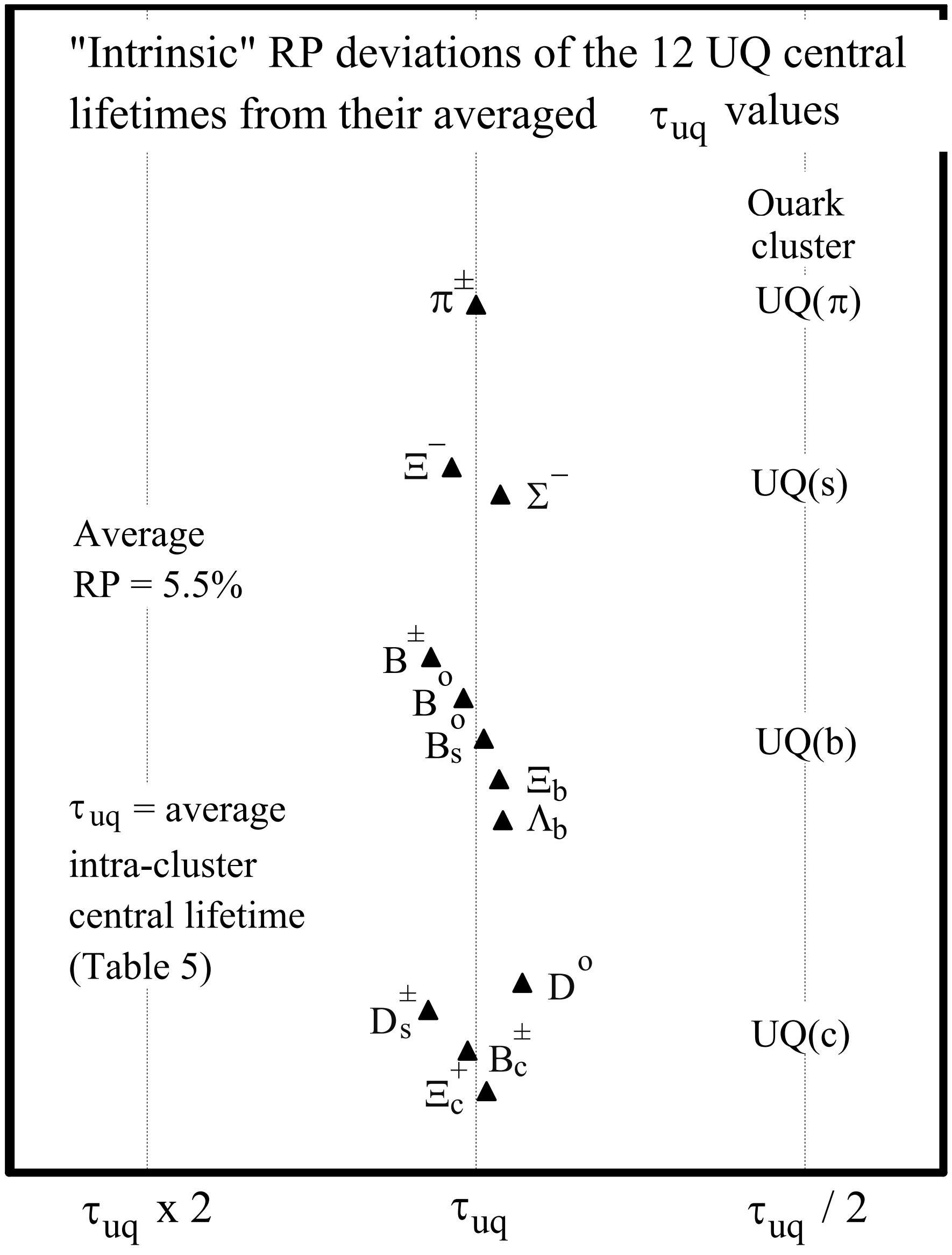

Fig. 10. The 12 central lifetimes of Fig. 6, shown here plotted as RP "random perturbation" deviations from the central lifetime $\tau_{uq}$ in each UQ lifetime cluster. The average RP deviation for the 11 lifetimes in these central-lifetime groups is 5.5% (with the $\pi^{\pm}$ reference lifetime excluded). These small RP deviations probably occur for a variety of reasons.

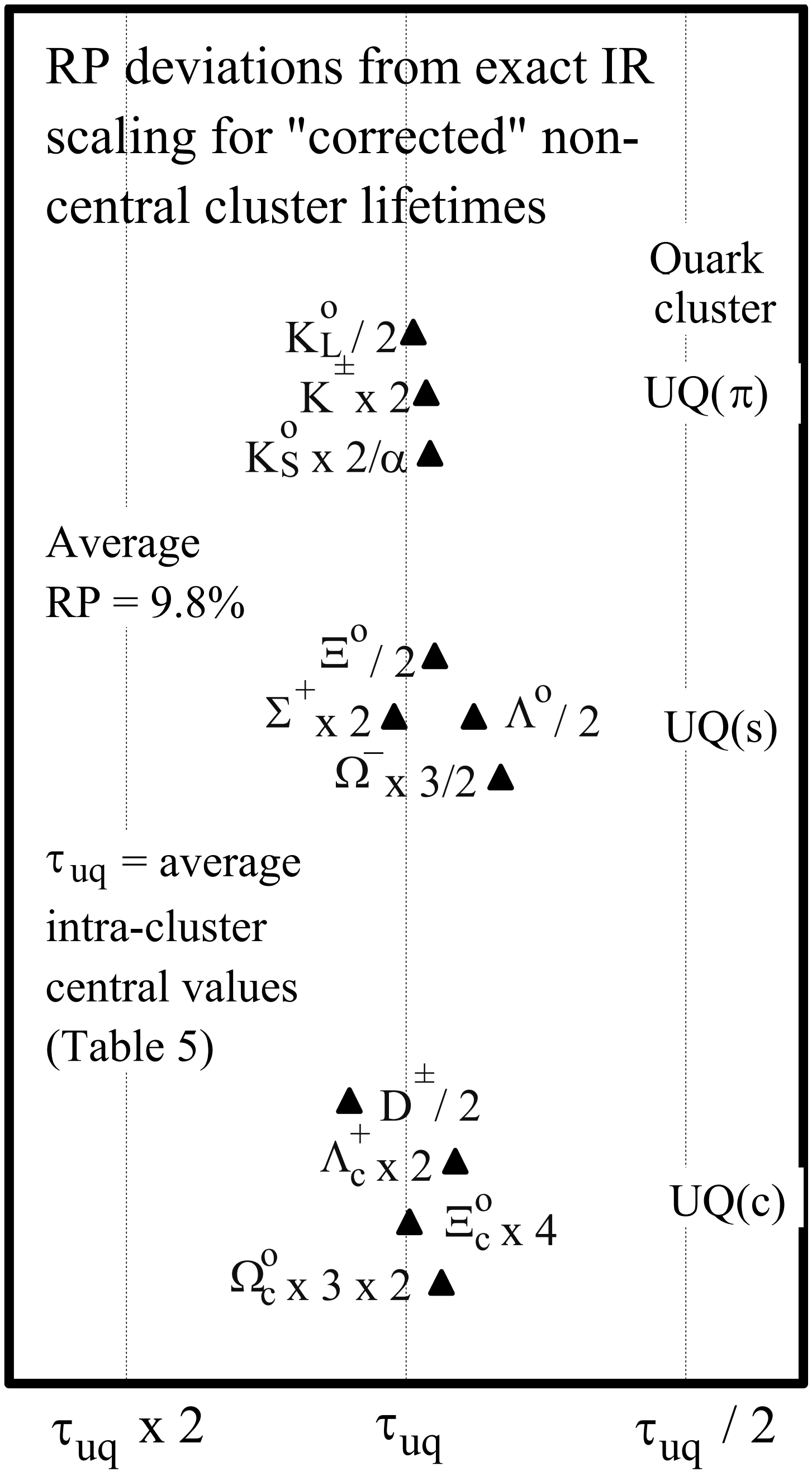

Fig. 11. The 11 displaced lifetimes of Fig. 6 after IR integer ratio "corrections" have been applied to move them into the central-lifetime region. They are displayed here as RP deviations from the central lifetime $\tau_{uq}$ in each UQ cluster. Their average (corrected) RP deviation is 9.8%, which is larger than the average RP deviation of 5.5% for the 12 central-lifetime particles displayed in Fig. 10, but of the same order of magnitude. Thus RP random perturbations are essentially the same for all of the particle lifetimes in a UQ cluster.

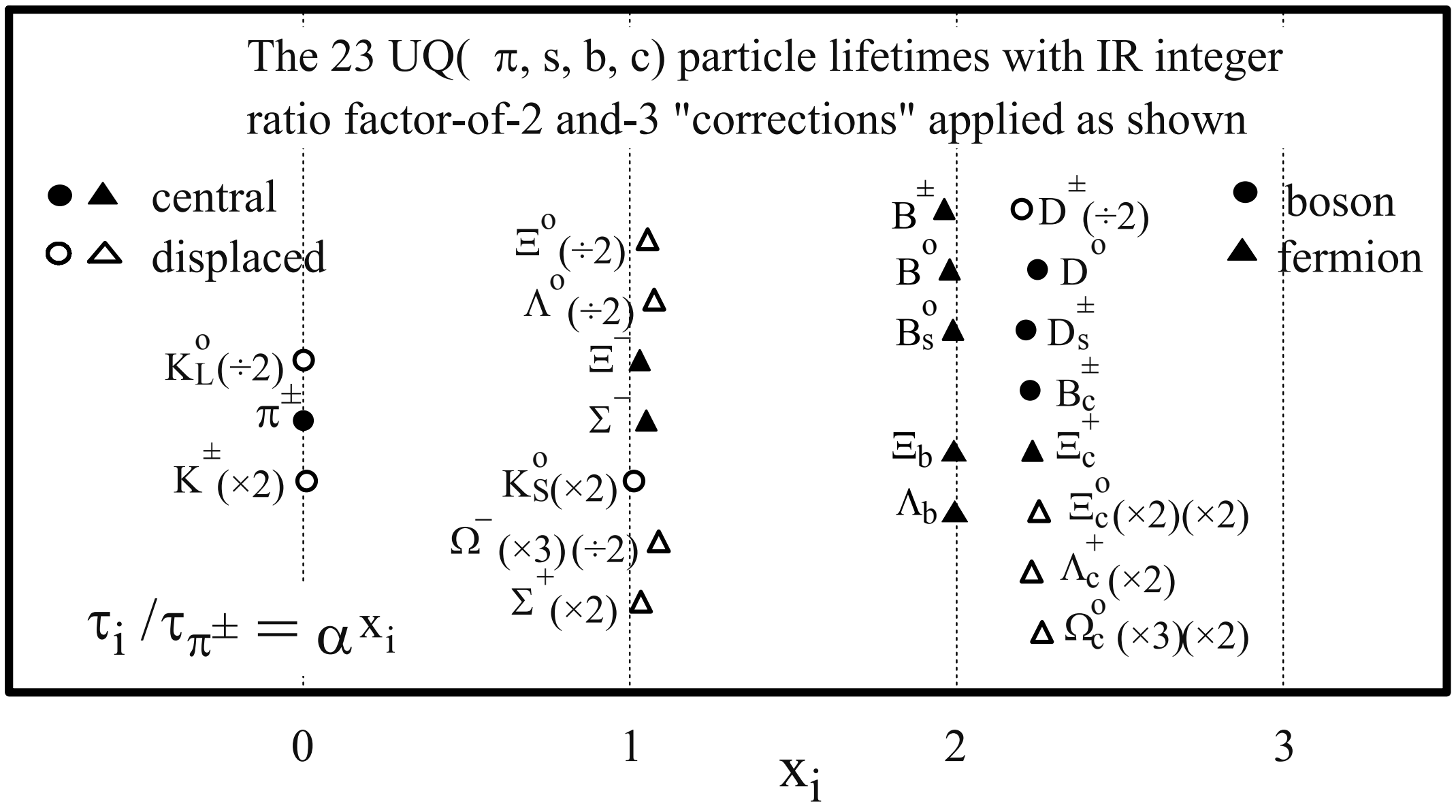

Fig. 12. The 23 UQ unpaired-quark particles of Fig. 6 after IR factors of 2 and 3 have been applied to move them into the central-lifetime regions of the UQ clusters. These empirical IR "corrections" enhance the sharpness of the lifetime flavor clusters. The effect of these corrections on the α-quantization of the clusters is displayed in Fig. 17.

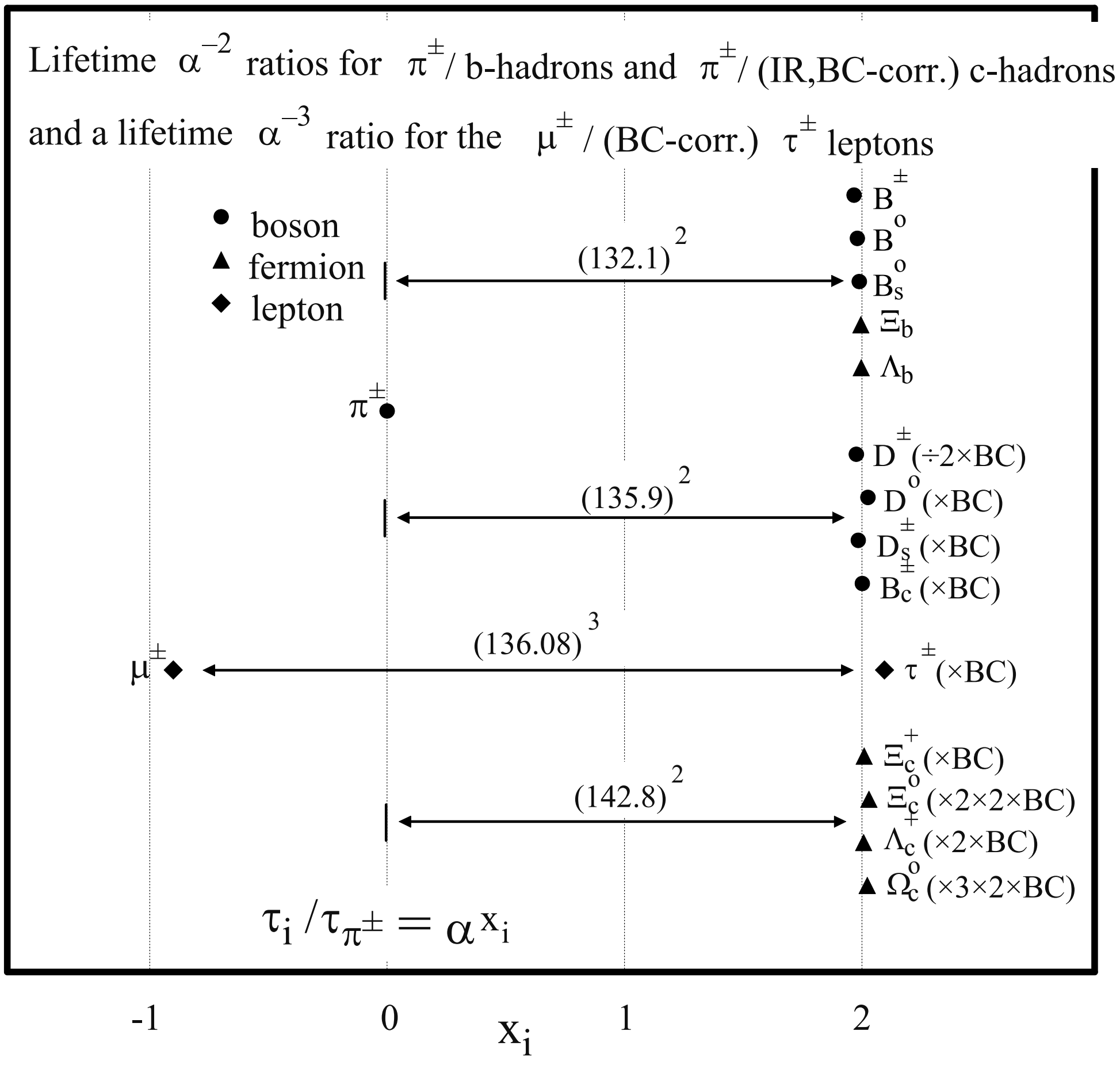

Fig. 13. The unpaired-quark UQ(b) meson states, which require no corrections, plotted together with the unpaired-quark UQ(c) meson states after IR = 2 and 3 and BC = 3.0 corrections are applied. With the $\pi^{\pm}$ as the reference lifetime, all of these states line up along the $x_i = 2$ α-spaced grid line (as $\alpha^{-2}$ lifetime ratios). Also shown is the τ lepton with the same BC = 3.0 correction applied. It now has a very precise $\alpha^3$ lifetime ratio with the μ lepton. The effect of changing the BC correction constant slightly is displayed in Figs. 18 and 19.

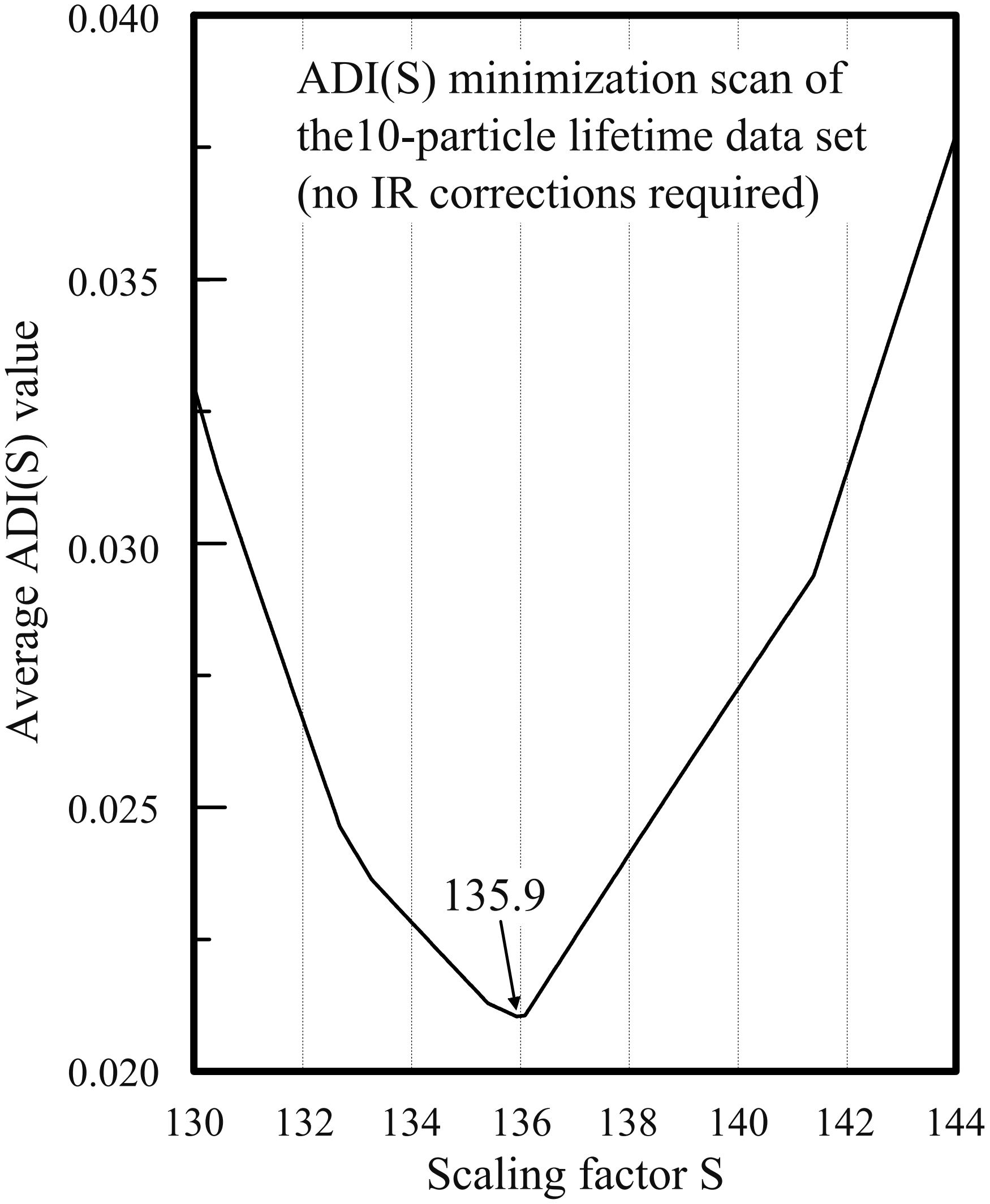

Fig. 14. The linear ADI (Absolute Deviation from an Integer) minimum for the 10-particle data set of Table 7 (which requires no IR or BC correction factors). The value $S_{min} = 135.9$ is close to the α-quantization scaling value $\alpha^{-1} \cong 137.0$. The value of ADI at $S_{min}$ is ADI ≅ 0.021, which compares to the value ADI ≅ 0.25 that is obtained for random scaling. An ADI value of 0.021 corresponds to an average lifetime deviation from exact α-scaling of about 15%.

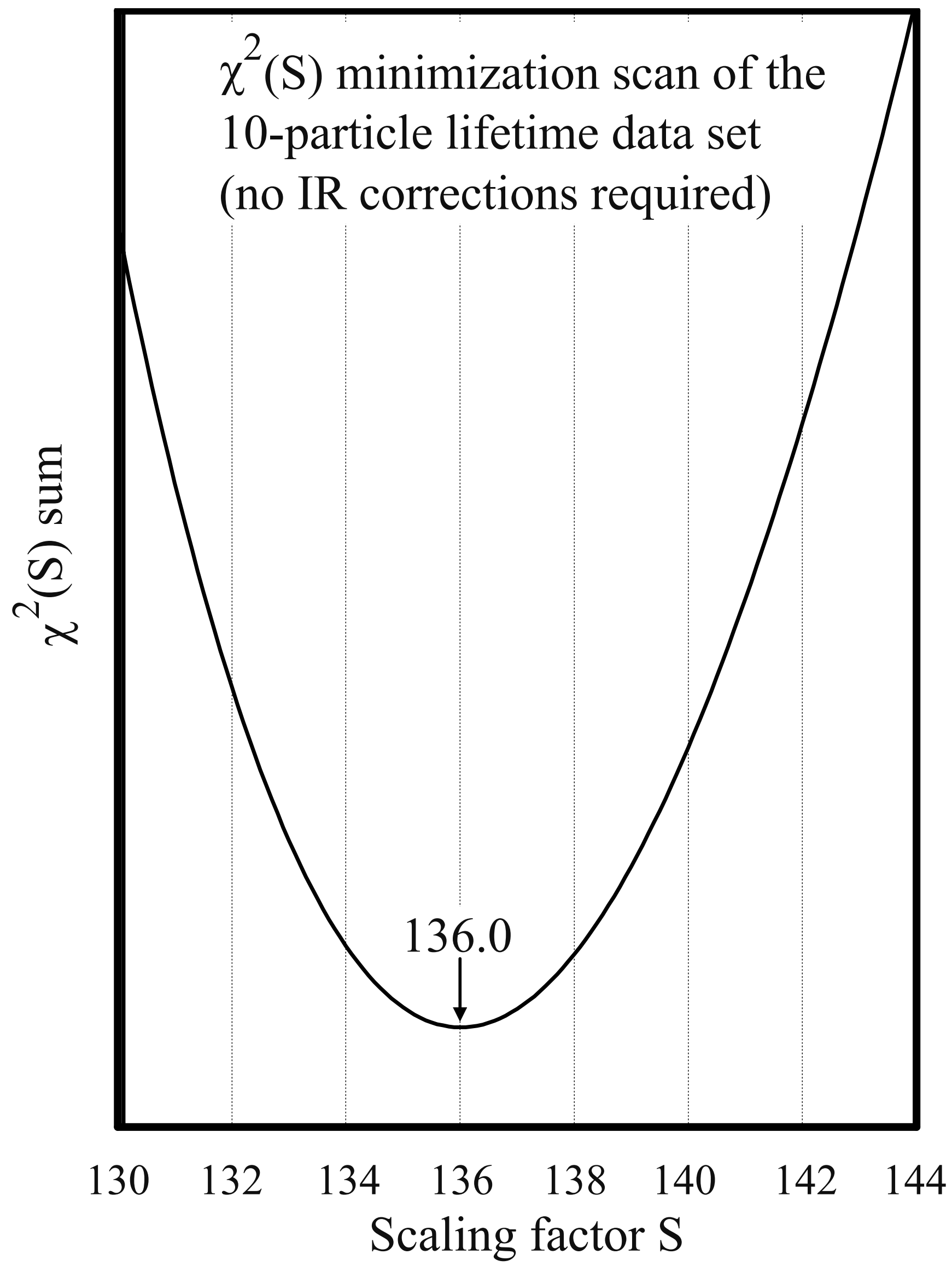

Fig. 15. The quadratic $\chi^2$ minimum for the 10 particle data set of Table 7. The value $S_{min}$ = 136.0 shown here closely matches the value $S_{min}$ = 135.9 that is obtained from the linear ADI minimization scan of Fig. 14.

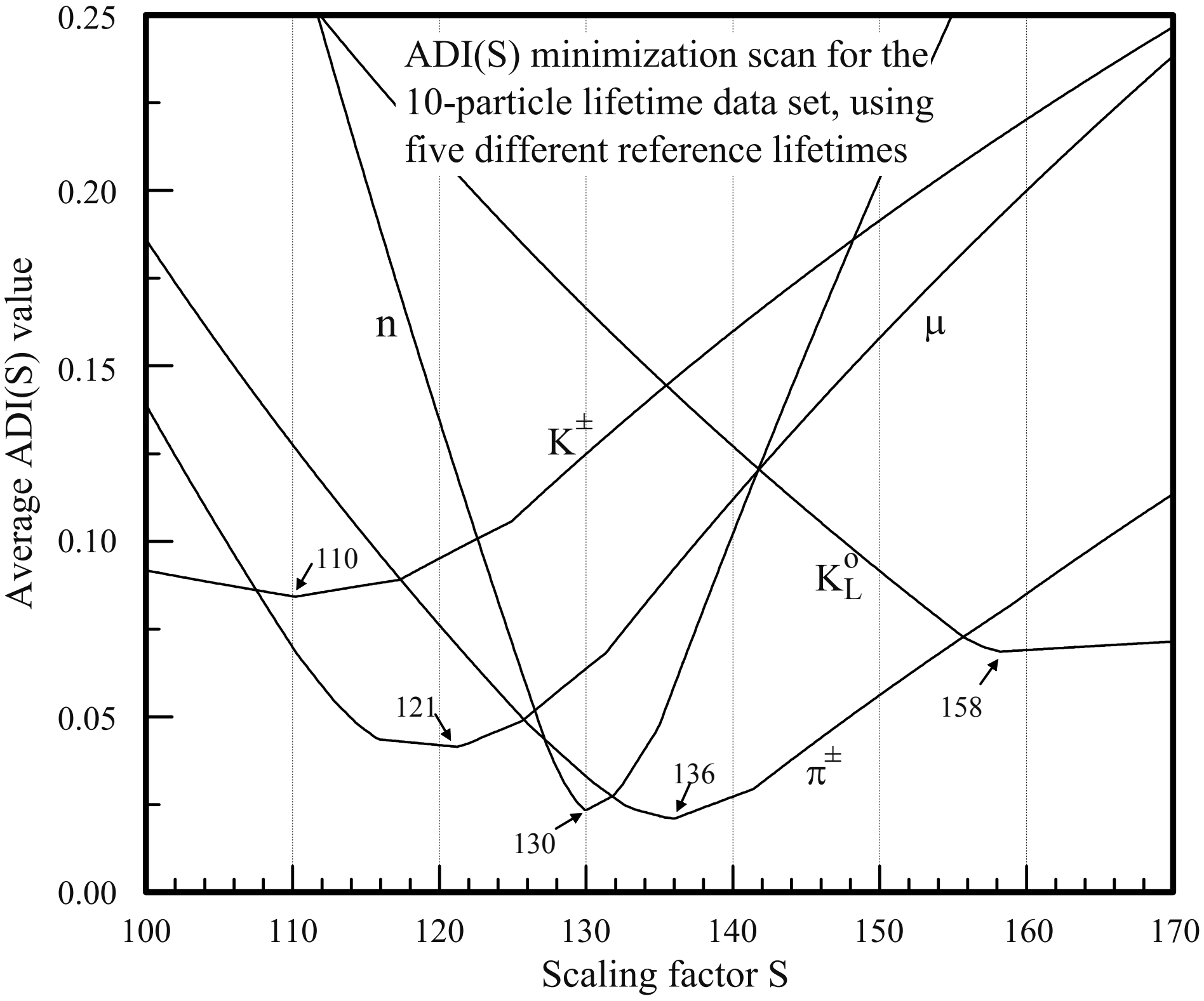

Fig. 16. ADI scans of the 10-particle data set, using α-quantized lifetime grids that are anchored on the five longest-lived metastable particles—the neutron, muon, $K_L^o$, $\pi^{\pm}$ and $K^{\pm}$. As can be seen, the α-grid that is anchored on the $\pi^{\pm}$ meson yields the lowest minimum value, ADI = 0.021, and it is the only one that has $S_{min} \cong \alpha^{-1} \cong 137$.

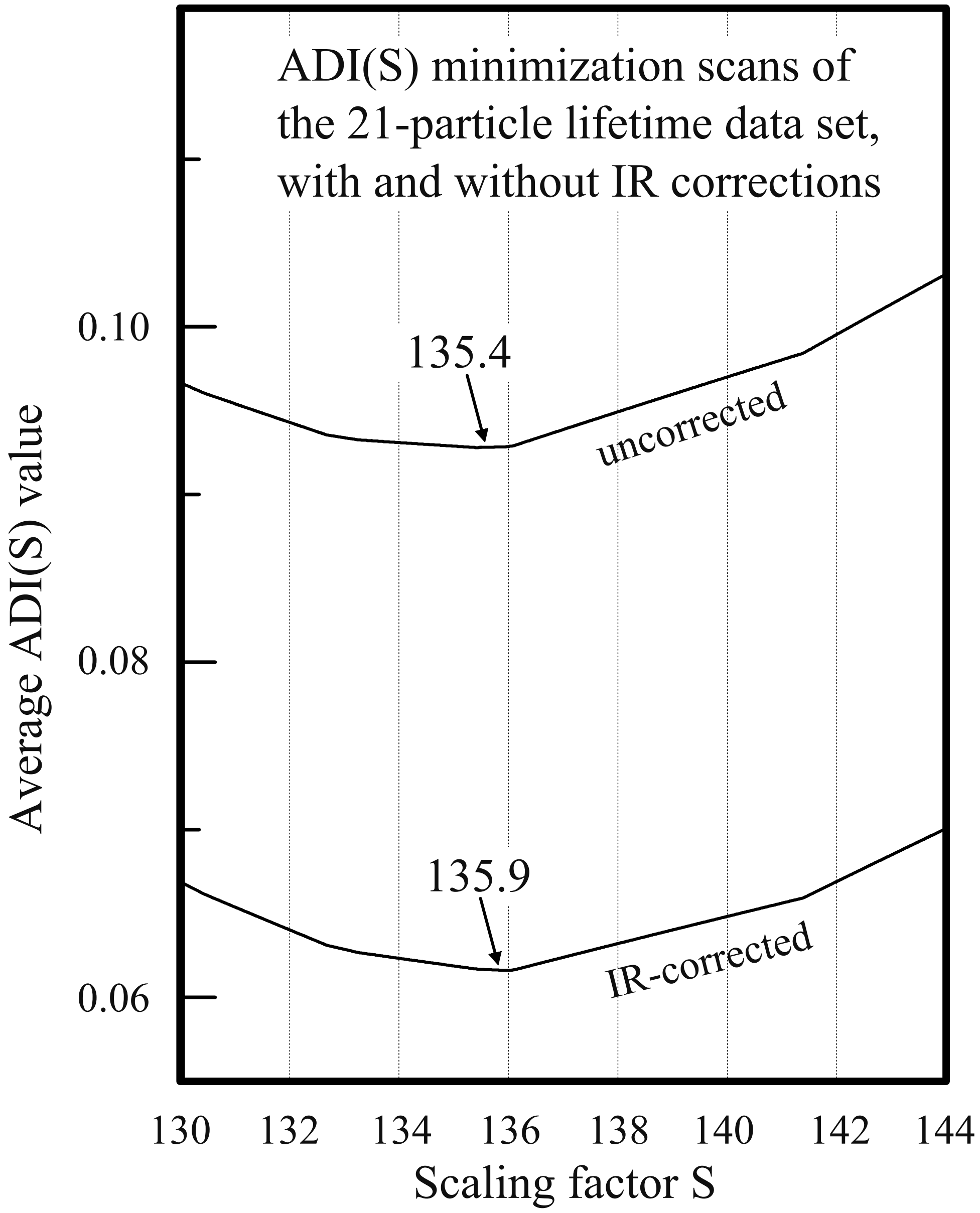

Fig. 17. ADI(S) scans of the 21-particle data set of Table 7, which includes particles that require IR integer ratio corrections. The effect of these IR corrections can be evaluated by comparing uncorrected and IR-corrected ADI(S) scans, as shown above. The IR corrections lower the value of ADI($S_{min}$) because of the increased lifetime clustering. They also shift $S_{min}$ towards the value $S_{min} \cong \alpha^{-1} \cong 137$. Thus these IR corrections slightly sharpen the $\alpha$-quantization of the metastable elementary particle lifetimes.

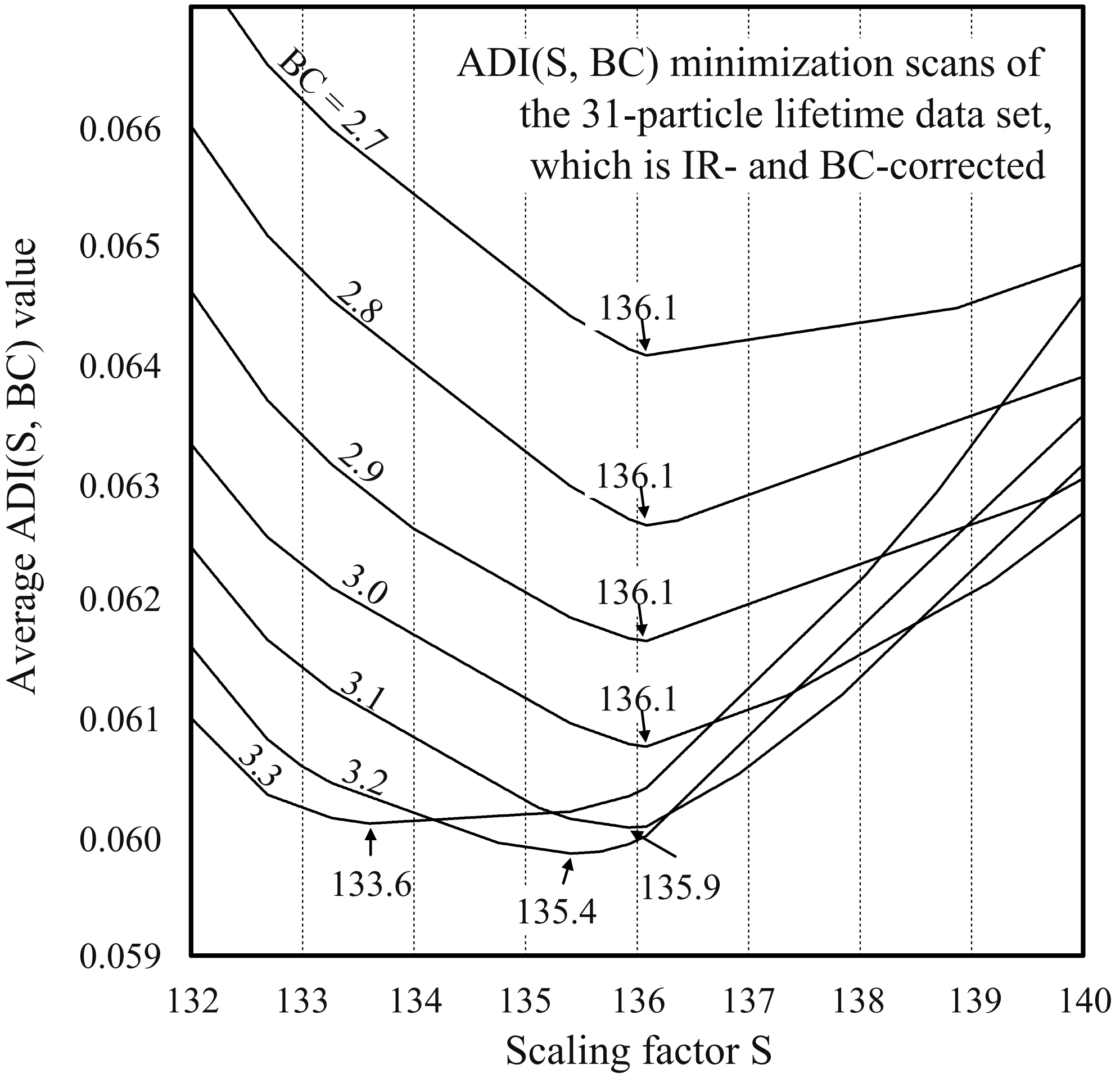

Fig. 18. ADI(S, BC) scans of the 31-particle data set of Table 7, using discrete values for the BC correction factor. As can be seen, the scaling factor $S_{min}$ stays constant at the value 136.1 as BC is increased from 2.7 up to 3.0. After that, $S_{min}$ decreases as a monotonic function of BC, which indicates that the *c*-quark lifetimes are being overcorrected. Hence these Fig. 18 ADI scans establish the upper limit BC < 3.1. This result is sharpened in Fig. 19.

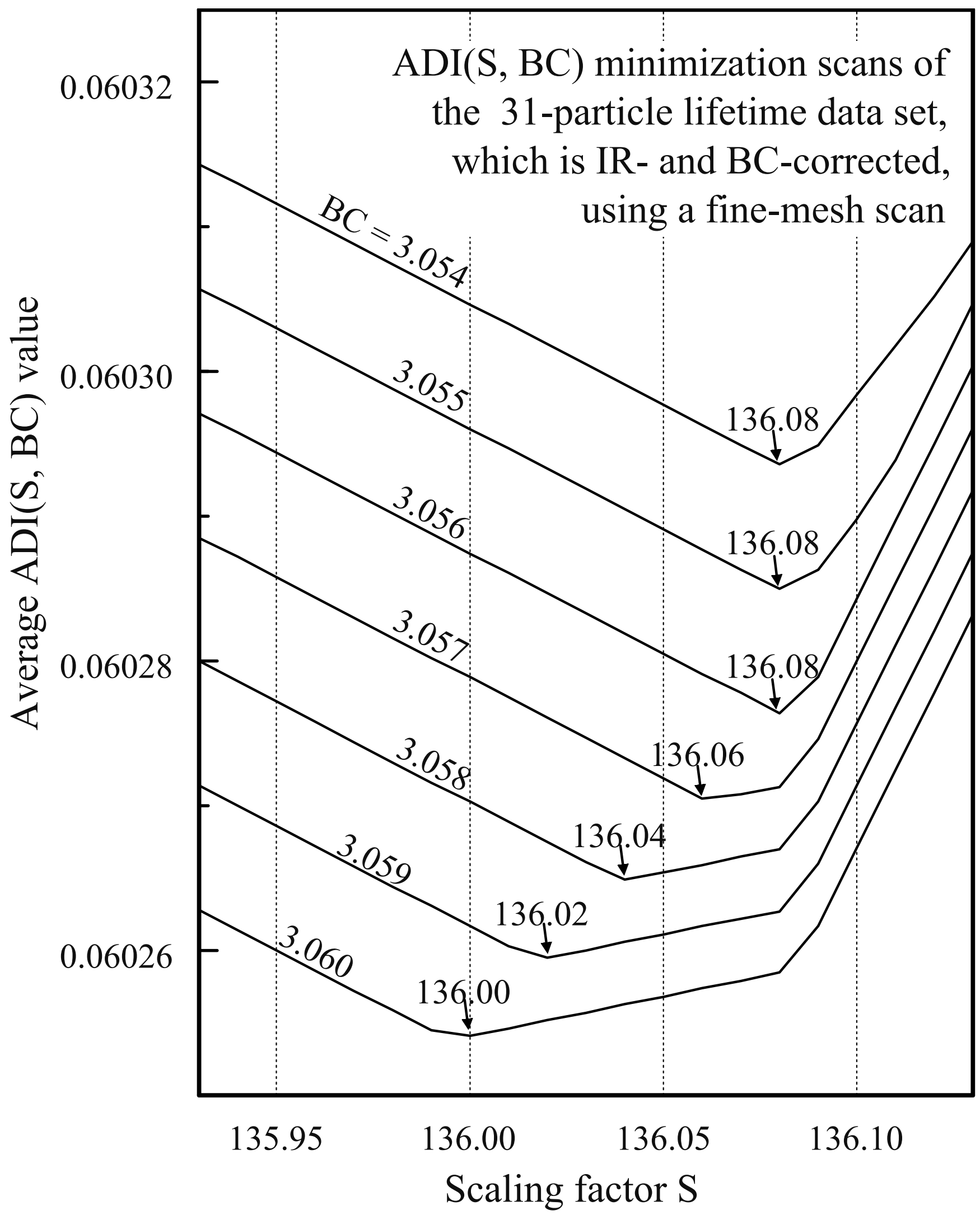

Fig. 19. The ADI(S, BC) scan of Fig. 18, carried out here with a very fine-mesh scan near the region of the allowable upper limit for the BC correction factor. The value of $S_{min}$ stays constant at the value 136.08 up to BC = 3.056, and then decreases monotonically at larger values. Hence these fine-mesh ADI scans establish the upper limit BC < 3.057.